\documentclass[aps,prl,preprint,groupedaddress]{revtex4}
\usepackage[dvips]{graphicx}
\begin{document}
\preprint{}
\title{Modelling Electron Spin Accumulation in a Metallic Nanoparticle}

\author{Y. G. Wei, C. E. Malec, D. Davidovi\'c }
\affiliation{Georgia Institute of Technology, Atlanta, GA 30332}

\date{\today}
\begin{abstract}
A model describing spin-polarized current via discrete energy
levels of a metallic nanoparticle, which has strongly asymmetric
tunnel contacts to two ferromagnetic leads, is presented.
 In absence of spin-relaxation, the model leads to a spin-accumulation in the
nanoparticle, a difference ($\Delta\mu$) between the chemical
potentials
 of spin-up and spin-down electrons, proportional to
the current and the Julliere's tunnel magnetoresistance. Taking
into account an energy dependent spin-relaxation rate $\Omega
(\omega)$, $\Delta\mu$ as a function of bias voltage ($V$)
exhibits a crossover from linear to a much weaker dependence, when
$|e|\Omega (\Delta\mu)$ equals the spin-polarized current through
the nanoparticle. Assuming that the spin-relaxation takes place
via electron-phonon emission and Elliot-Yafet mechanism, the model
leads to a crossover from linear to $V^{1/5}$ dependence. The
crossover explains recent measurements of the saturation of the
spin-polarized current with $V$ in Aluminum nanoparticles, and
leads to the spin-relaxation rate of $\approx 1.6 MHz$ in an
Aluminum nanoparticle of diameter $6nm$, for a transition with an
energy difference of one level spacing.
\end{abstract}

\maketitle

\section{introduction}

Spin-dependent electron transport through nanometer scale
structures has attracted an increased interest
recently.~\cite{seneor,ernult} In general, short spin-diffusion
lengths in metals makes it necessary to investigate spin-dependent
transport in micron-scale metallic
structures.~\cite{johnson,jedema} More recently, spin-dependent
electron transport has been investigated in single nanometer-scale
metallic particles,~\cite{bernard,wei} including spin-polarized
transport via discrete electronic energy levels of a
nanoparticle.~\cite{wei}

\begin{figure}
\includegraphics[width=0.9\textwidth]{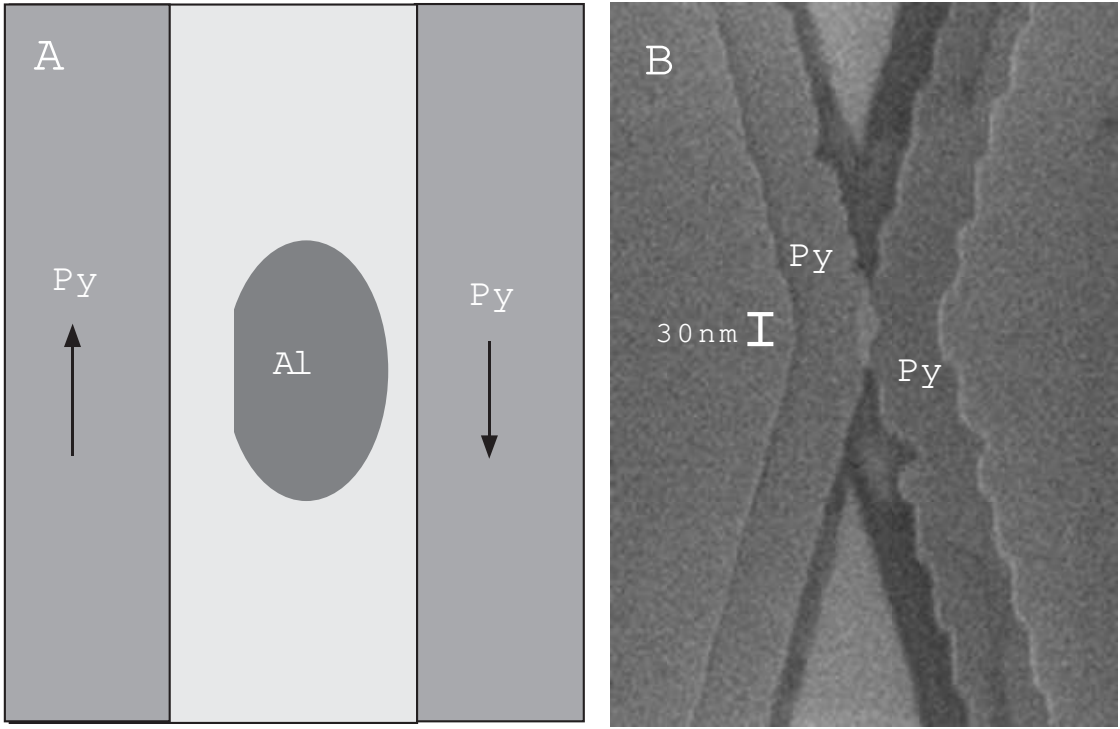}
\caption{A: Sketch of sample cross-section in the antiparallel
magnetization configuration of the leads. B. Scanning Electron
micrograph of a typical device. The tunneling junction with
embedded nanoparticles is located inside the overlap of two Py
leads.\label{device}}
\end{figure}

In this paper, we develop a model to explain a previously reported
experiment designed to detect spin polarized currents in a normal
Aluminum nanoparticle connected to ferromagnetic leads by weak
tunnel barriers.~\cite{wei}  These experiments were carried out on
lithographically defined tunnel junctions as featured in fig.
~\ref{device}.  In principle, spin polarized current can be
determined by analyzing the
$TMR=(I_{\uparrow\uparrow}-I_{\uparrow\downarrow})/I_{\uparrow\uparrow}$,
where $I_{\uparrow\uparrow}$ and $I_{\uparrow\downarrow}$ are the
currents through the nanoparticle in the parallel and antiparallel
magnetization configurations.

It is assumed here that the difference between
$I_{\uparrow\uparrow}$ and $I_{\uparrow\downarrow}$ arises from
the differences in spin-up and spin-down densities of states in
the leads. In sequential electron tunneling via the nanoparticle,
spin-polarized current is a consequence of spin accumulation, a
difference in the chemical potentials of the spin-up and spin-down
electrons in the nanoparticle, caused by the tunnel electric
current and the spin-polarized densities of states.
~\cite{barnas,majumdar,barnas1,brataas,brataas1,korotkov1,barnas3,barnas2,imamura2,kuo,weymann1,
brataas2,wetzels,weymann2,braun,braun1}  Spin-accumulation is
found only in the antiparallel magnetization configuration,
because only in that case the ratios of the tunnel-in and
tunnel-out resistances of the spin-up and the spin down electrons
are different. A necessary condition for spin-accumulation is that
the electron spin be conserved during the sequential transport
process.

There are several compelling reasons to study TMR in nanometer
scale particles at temperatures where discrete energy levels can
be resolved.  One is the magneto-coulomb effect
~\cite{vandermolen,kuo} which can give a strong TMR signal even
without spin-accumulation in the nanoparticle.  In the regime of
well resolved energy levels the contribution arising from
spin-polarized current and spin-accumulation can be separated from
the contribution arising from the chemical potential
shifts.~\cite{wei} In particular, if the leads chemical potentials
vary in a range that corresponds to a nanoparticle energy range
that is in between two successive discrete energy levels, then
electron current through the nanoparticle will be insensitive to
changes in the chemical potentials. In terms of the I-V curve,
which exhibits step like increases at bias voltages corresponding
to discrete energy levels, $TMR$ must be measured between the
current steps, at voltages where the current as a function of bias
voltage is constant.

Another reason is that nanoparticles of this size exhibit
extraordinarily weak spin-orbit coupling compared to bulk.  In
this regime, the stationary electron wavefunctions are slightly
perturbed spinors.~\cite{matveev,brouwer} As a result,  the spins
of electrons injected from a ferromagnet into the nanoparticle
have exceptionally long life-times. This remarkable regime of
spin-polarized electron transport via metallic nanoparticles has
hardly been explored experimentally.~\cite{bernard,wei} By
contrast, most measurements of electron spin-injection and
accumulation in metals~\cite{johnson,jedema} were obtained in the
regime of strong spin-orbit scattering, where any nonequilibrium
spin-population exhibits exponential time-decay.

The main result of the prior report~\cite{wei} was that the
spin-polarized current through the nanoparticle saturates quickly
as a function of bias voltage, for tunnel resistances in the
$G\Omega$ range. In that case the saturation is reached typically
around the second or third energy level of the nanoparticle.

The saturation effect was explained by a rapid increase of the
spin-relaxation time with the nanoparticle excitation energy. By
this interpretation, the spin-polarized current through the
nanoparticle is carried only via the ground state and the few
lowest energy excited states of the nanoparticle, while highly
excited spin-polarized states relax faster than the average
sequential electron tunnel process. We conjectured that at the
saturation voltage, the relaxation time of the highest singly
occupied energy level of the nanoparticle is comparable to the
electron tunnel rate. The corresponding spin relaxation time is in
the microsecond range. In comparison, the spin relaxation time in
Aluminum thin films with a similar mean free path would be five
orders of magnitude shorter.

Our measurement of the spin-relaxation time was somewhat indirect,
because theoretical literature prior to our work had not predicted
any saturation of the spin-polarized current with bias
voltage.~\cite{barnas,majumdar,barnas1,brataas,brataas1,korotkov1,barnas3,barnas2,imamura2,kuo,weymann1,
brataas2,wetzels,weymann2,braun,braun1} Consequently our
explanation of the saturation was qualitative. The goal of this
paper is to obtain a model of spin-polarized electron transport
through a metallic nanoparticle to explain our observations. We
use a method for calculating current via energy levels of the
nanoparticle based on rate equations, following Ref.~\cite{bonet}.
The tunneling regime in our devices is different from that used in
the theoretical studies, because the tunnel resistances in our
junctions are highly asymmetric and spin-relaxation rate has
strong energy
dependence.~\cite{barnas,majumdar,barnas1,brataas,brataas1,korotkov1,barnas3,barnas2,imamura2,kuo,weymann1,
brataas2,wetzels,weymann2,braun,braun1}

Other experimental work on arrays and single nanoparticles did not
find any saturation of the spin-polarized current with bias
voltage.~\cite{yakushiji,Yakushiji2002,Yamane2004,zhang1,Yakushiji2007,Yakushiji2007a,bernard}
These experiments do not  measure TMR in the regime of well
resolved energy levels, we believe that this is a critical
measurement to separate the contributions to $TMR$ from the
chemical potential shifts.
The analysis of $TMR$ experiments in Ref.~\cite{seneor}
uses an energy independent spin-relaxation time.  It is possible,
however; that the energy dependence of the spin-relaxation rate can be quite
strong.~\cite{yafet}  We show in this paper that the effect of
the energy dependence on spin-polarized current is significant.

Our model is valid within a specific experimental regime, outlined in Sec.~\ref{rmv},
but it is an easily analyzable and experimentally relevant regime that merits
consideration as a large number of samples fall under this parameter range.

In Sec.~\ref{eso} we review the effects of spin-orbit scattering
in metallic nanoparticles, in the context of spin-injection and
detection. In Sec.~\ref{ermn} we discuss various energy-relaxation
rates in the nanoparticle. The criteria of the model validity are
listed on Sec.~\ref{rmv}.  We then calculate the probability
distribution of the many-electron states in the nanoparticle in
Sec.~\ref{cspd} and use this to calculate a TMR versus bias
voltage curve that can be fit to experimental data in
Sec.~\ref{fit}

\section{Effects of spin-orbit interaction on spin-polarized current through a nanoparticle.}
\label{eso}

In a metallic nanoparticle, the stationary electronic
wavefunctions
 in zero applied magnetic field are two fold degenerate
and form Kramers doublets:
 \[
|\uparrow>'=u(\vec{r})|\uparrow>+v(\vec{r})|\downarrow>,
\]
\[
|\downarrow>'=u^\star(\vec{r})|\downarrow>-v^\star(\vec{r})|\uparrow>.
\]
The mixing between the spin-up and the spin-down components is
caused by the spin-orbit interaction. In a magnetic field, the
degeneracy is lifted by the Zeeman effect. If the field is applied
in direction corresponding to $|\uparrow>$,  the g-factor is
$g=2-4\int d^3\vec r|v|^2<2$.

The effects of the spin-orbit interactions on the wavefunctions
are described by a dimensionless parameter $\lambda$
\begin{equation}
\lambda=\sqrt{\frac{\pi\hbar\nu_{SO}}{\delta}}, \label{soregime}
\end{equation}
where $\nu_{SO}$ is the spin-orbit scattering rate and $\delta$ is
the average spacing between successive Kramers doublets in the
particle.~\cite{matveev,brouwer,adam} The effects of spin orbit
scattering are weak if $\lambda\ll 1$. In that case,
$v(\vec{r})\approx 0$, $u(\vec{r})$ is real, and the wavefunctions
$|\uparrow>'$ and $|\downarrow>'$ have well defined spin. In the
opposite limit, $\lambda\gg 1$, the spin-orbit scattering is
strong. In that case, $\int d^3\vec r |u|^2\approx \int d^3\vec r
|v|^2\approx 1/2$ and the wavefunctions have uncertain spin.

Consider an electron with spin-up injected by tunneling from a
ferromagnet into the nanoparticle. If the tunnel process is
instantaneous, the electron will have a well defined spin
$|\uparrow>$ immediately after tunneling. In the regime of weak
spin-orbit scattering, where $v\approx 0$, the initial state has a
much larger overlap with a wavefunction $|\uparrow>'$ than with
the corresponding wavefunction $|\downarrow>'$. As a result, the
spin of the added electron will remain well defined, in principle
indefinitely long, barring any coupling between the nanoparticle
and the environment. In that case, the detection of the injected
spin can be performed at any time after the injection.

By contrast, in the regime of strong spin-orbit scattering, the
initial state overlaps with both $|\uparrow>'$ and
$|\downarrow>'$, nearly equally. In that regime, the spin of the
added electron becomes uncertain after a time $\sim
\tau_{SO}=1/\nu_{SO}$, so there is a time limit $\sim\tau_{SO}$
for spin detection.

In normal metals, there is a scaling between the spin-conserving
electron scattering rate and the corresponding spin-flip electron
scattering rate,~\cite{yafet,elliot}
\begin{equation}
\nu_{SO} (\omega)=\alpha\nu (\omega). \label{eliot}
\end{equation}
This is known as the Elliot-Yafet relation. $\omega$ is the energy
difference between the initial and final state. For elastic
scattering, $\omega=0$ and $\nu_{SO} (0)=\nu_{SO}$. The scaling
parameter $\alpha$ depends on the spin-orbit scattering and the
band structure. In Aluminum, $\alpha\approx 10^{-4}$ is larger
than anticipated from the spin-orbit interaction, because of the
hot-spots for spin scattering in the band
structure.~\cite{fabian,fabian2}

The Elliot-Yafet relation was confirmed in bulk metals by the
conduction-electron spin resonance experiments (CESR). In
particular, the temperature dependence of the width of the
spin-resonance line, which is proportional to $\nu_{SO}(k_BT)$,
follows the temperature dependence of the resistivity, which is
proportional to the momentum relaxation rate, in agreement with
Eq.~\ref{eliot}. The relation has also been confirmed more
recently in mesoscopic metallic samples, by the spin-injection and
detection experiments.~\cite{jedema,jedema1,jedema3} Both CESR and
spin-injection and detection experiments measure the time decay of
a nonequilibrium spin population in the regime of strong
spin-orbit scattering.

In metallic nanoparticles, the validity of the Elliot-Yafet
relation was confirmed experimentally by energy level
spectroscopy, in the regimes of weak and moderate spin-orbit
scattering.~\cite{petta} $\nu_{SO}$ can be measured directly from
the magnetic field dependence of the energy levels.~\cite{adam}
Petta {\it et al.}~\cite{petta} found that in Cu, Ag, and Au nanoparticles,
$\nu_{SO}\approx \alpha v_F/D$, where $v_F$ is the Fermi velocity
and $D$ is the nanoparticle diameter, and $\alpha$ is close to the
bulk value, within an order of magnitude. This confirms the
Elliot-Yafet relation because $v_F/D$ is equal to the elastic
scattering rate, assuming a ballistic nanoparticle. More recently,
the Elliot-Yafet relation was confirmed in Al nanoparticles as
well.~\cite{wei}

Substituting the level spacing and the Elliot-Yafet relation into
Eq.~\ref{soregime}, we find that there is a characteristic
nanoparticle size
\begin{equation}
D^\star=\lambda_F/\sqrt{\alpha}. \label{dstar}
\end{equation}
Spin-orbit scattering will be weak if $D\ll D^\star$ and
spin-orbit scattering will be strong if $D\gg D^\star$. $D^\star$
is a material dependent microscopic
parameter. \footnote{Similarly, in a diffusive nanoparticle we
find $D^\star_{dif}\sim(\lambda_F^2\l/\alpha)^{1/3}$, where $l$ is
the mean free path.} In Aluminum, $D^\star\approx
10^2\lambda_F\sim 10nm$.

It follows that both mesoscopic and macroscopic metallic samples
are in the regime of strong spin-orbit scattering. Only if the
nanoparticle diameter is less than about 10nm the spin-orbit
scattering becomes weak. In samples much larger than $D^\star$,
measuring the discrete electron energy levels and the electron
spin polarization are incompatible, and the spin-polarized current
via resolved energy levels must be negligibly small. We are not
aware of any theory that explicitly takes into account the effect
of spin-mixing in the Kramers doublets on $TMR$ of ferromagnetic
single-electron transistors, e.g. that calculates $TMR$ versus
$\lambda$. Such a theory would be important because it would set
the limits of observability of spin-polarized current through
single electron transistors.

In samples much larger than $D^\star$, one can still study the
time decay of injected electron spins at a time scale much shorter
than the time necessary to resolve discrete energy levels (the
Heisenberg time $\hbar/\delta$), as demonstrated by the CESR
experiments and spin-injection and detection in mesoscopic metals.
In further discussion, we assume $D\ll D^\star$. In that case, the
spin up band in the ferromagnet is tunnel coupled  to the
nanoparticle wavefunctions with spin-up only.

\section{Energy Relaxation in Metallic Nanoparticles}
\label{ermn}

In this work we study spin-polarized electron current via discrete
energy levels of metallic nanoparticles, in a regime where the
tunnel rate is much smaller than the spin-conserving energy
relaxation rate.  The tunnel rate can still be larger than the
spin-flip relaxation rate, so electrons in the nanoparticle are
not in equilibrium.

\begin{figure}
\includegraphics[width=0.9\textwidth]{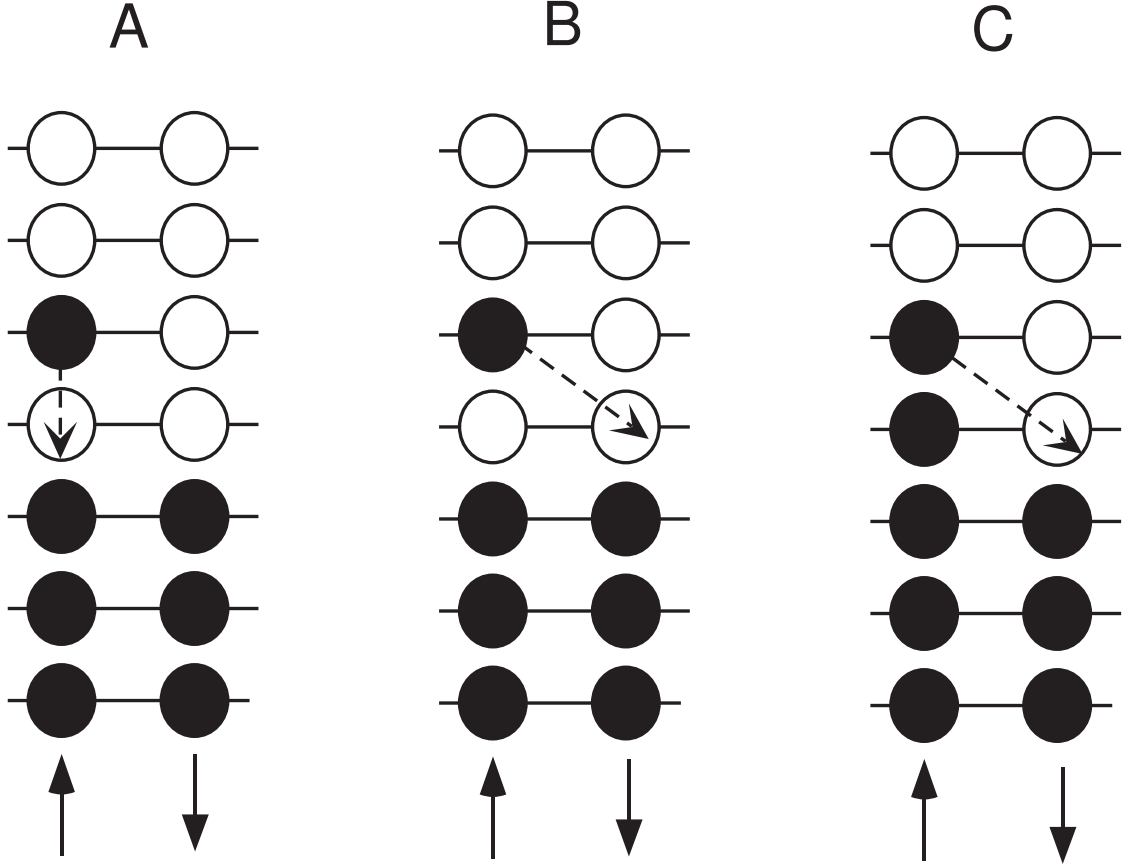}
\caption{A: Spin-conserving energy relaxation process.  B.
Spin-flip energy relaxation process. C: Electronic state with the
relaxation time equal to the spin-flip relaxation
time.~\label{excite}}
\end{figure}

Consider a nanoparticle with an electron added at energy
$\omega=\delta $, as shown in Fig.~\ref{excite}. The nanoparticle
is in an excited state and it can relax its energy internally. It
was shown that the dominant relaxation process is electron-phonon
interaction.~\cite{agam} One possible path for this relaxation is
spin-conserving, as shown in Fig.~\ref{excite}-A. The
spin-conserving relaxation rate via phonon emission have been
estimated by Agam et al.~\cite{agam}
\begin{equation}
{\nu_{e-ph}(\omega)}=\left(\frac{2}{3}E_F\right)^2\frac{\omega^3\tau_e\delta}{2\rho\hbar^5
v_s^5}, \label{eph}
\end{equation}
 where $E_F=11.7 eV$ is the Fermi
energy, $\omega$ is the energy difference between the initial and
the final state, $\tau_e$ is the elastic scattering relaxation time, $\rho=2.7 g/cm^3$ is the ion-mass density, and
$v_s=6420 m/s$ is the sound velocity. This formula is equivalent
to
\[
\hbar\nu_{e-ph}(\omega)\sim
\frac{1}{g}E_F\left(\frac{\omega}{\omega_D}\right)^3,
\]
within a prefactor of order 1, where $\omega_D$ is the Debye
energy and $g$ is the dimensionless conductance of the
nanoparticle, $g=E_{Th}/\delta$, and $E_{Th}=\hbar v_F/D$ is the
Thouless energy. This relation is similar to the formula for bulk
electron-phonon scattering rate, except that there is a prefactor
$1/g$,~\cite{agam} which originates from the chaotic nature of the
electron wavefunctions.

We estimate $\nu_{e-ph}(\delta)\sim GHz$ in nanoparticle with
diameter $5nm$, in agreement with experiment.~\cite{ralph} All our
samples have tunnel rates significantly smaller than this
relaxation rate, so, the nanoparticle can be considered to be
relaxed toward the lowest energy state accessible via
spin-conserving transitions.

The nanoparticle can also relax through a transition depicted in
Fig.~\ref{excite}-B, with some rate $\nu^{SF}(\omega)$,
$\omega=\delta$. This spin-flip transition involves coupling
between electrons and the environment, which may be the phonon
bath or the bath of nuclear spins. In zero applied magnetic field,
the electron transitions shown in Figs.~\ref{excite}-A and B have
the same energy difference between the initial and the final
states.
 We expect that the
transition rate for the spin-flip relaxation process is much
smaller than the corresponding spin-conserving relaxation rate,
because in the spin-flip process there must be a transfer of
angular momentum into the environment.

For example, if the environment is the ion lattice, then the
transfer is governed by the spin-orbit interaction, which is much
smaller than the the electrostatic interaction that governs the
spin-conserving transitions. Thus the spin-flip probability caused
by the spin-orbit interaction during a phonon emission process is
very small. It is reasonable to expect that the Elliot-Yafet
scaling is valid for the transitions between chaotic
wavefunctions, so that
$\nu^{SF}(\omega)=\alpha\nu_{e-ph}(\omega)$, but we are not aware
of any theoretical calculation of the spin relaxation rate in
metallic nanoparticles.

Spin-flip electron transition rates between discrete levels were
obtained theoretically for semiconducting quantum
dots.~\cite{khaetskii,khaetskii1} The theoretical calculations
were in good agreement with the experimental results in GaAs
quantum dots.~\cite{hanson} The theory of spin-relaxation in
semiconducting quantum dots is of no use for metallic
nanoparticles, because the mechanisms of spin-orbit interaction in
metals and semiconductors are very different.~\cite{Zutic2004}

It would be very difficult to measure the transition rate depicted
in Fig.~\ref{excite}-B, because the relaxation process in
Fig.~\ref{excite}-A competes with that in Fig.~\ref{excite}-B. We
are able to determine the spin relaxation rates in tunneling
measurements because it is possible to trap the nanoparticle into
a state shown in Fig.~\ref{excite}-C. In that case the
spin-conserving energy relaxation is forbidden by the Pauli
principle and the relaxation rate is equal to the spin relaxation
rate.

\section{Region of model validity}
\label{rmv}

In this paper the tunnel current via discrete energy levels of a
metallic nanoparticle is calculated using the methods outlined in
Ref.~\cite{bonet} We consider spin polarized tunneling in the
regime closest to our experiments. The following conditions define
that regime.

1. We assume that the tunnel junctions in our samples are highly
asymmetric in resistance. The asymmetry arises because the tunnel
junctions are fabricated using conventional lithography and
evaporation. A weak non-uniformity in the tunnel junction
thickness leads to large asymmetry in the tunnel
resistance.~\cite{wei}

In these asymmetric samples, the current through the nanoparticle
is limited by the tunnel rate through the high resistance
junction. The resistance of the low resistance junctions is less
directly related to the current and it can be obtained by
comparing the amplitudes of Zeeman split energy levels at positive
and negative bias voltage. That procedure has been described in
detail in some special cases in Ref.~\cite{bonet} Even though our
regime is quite different from those special cases, they are still
indicative of the procedure to determine the resistance ratio.
Applying the general formalism of Ref.~\cite{bonet} to our regime
we estimate the resistance ratio to be about 25 in sample 1. The
results of our model do not depend on the resistance ratio, as
long as the electron discharge rate is much larger than the
electron tunnel in rate.

2. We calculate the current in the regime when the first tunnel
step is across the high resistance junction. This step is followed
by an electron discharge via the low-resistance junction. This
regime is relatively easy to analyze because electron discharge is
fast and so the charging effects do not influence the current
significantly.~\cite{averin1,bonet} In this case, the nanoparticle
spends most of the time waiting for an electron to tunnel in.

3. We assume $E_C\gg \delta$, where $E_C$ is the charging energy.
This assumption is generally valid in metallic
nanoparticles.~\cite{ralph}

4. We assume that the coulomb gap in the I-V curve is much larger
than the level spacing, this requires not only that condition 3 is met,
but that the background charge $q_0$
is not too close to $(n+1/2)e$, where $n$ is an integer. In that
case the number of energy levels participating in electron
transport is always $\gg 1$. Even if an electron tunnels into the
lowest unoccupied single-electron energy level of the
nanoparticle, there will still be a large number of occupied
single-electron energy levels of the nanoparticle that can
discharge an electron.

5. We assume that the number of electrons on the nanoparticle
before tunneling in is even. The calculation for the odd case is
very similar to that for the even case and will not be discussed
here.

6. We assume that the tunnel rates between the single-electron
states $i$ in the nanoparticle and the leads are much smaller than
the spin-conserving energy relaxation rates, which are $>GHz$. In
the sample selected for this paper, the tunnel rate across the
high resistance junction is in the $MHz$ range.
\\\indent If the spin-relaxation process is taking place and there is a
large asymmetry in junction resistances, an important question is
which one of the two tunnel rates limits the $TMR$. We will assume
that the left junction has higher resistance. It may be tempting
to assume that $TMR$ will be highly asymmetric with bias voltage
if $\Gamma^{L}\ll \nu^{SF}(\omega)\ll \Gamma^R$, where
$\Gamma^{L}$ and $\Gamma^{R}$ are the electron tunnel rates
between the energy levels of the nanoparticle and the left and the
right leads, respectively, because, if a spin-polarized electron
tunnels in via the high resistance junction, it will tunnel out
via the low resistance junction before spin-relaxation process
takes place, and, if the direction of the current is reversed, the
order of tunneling will be reversed and the spin-relaxation will
take place before tunneling out. A similar situation is found in
measurements of the energy spectra in samples where $\Gamma^{L}\ll
\nu_{e-ph}(\omega)\ll \Gamma^R$, where $\nu_{e-ph} (\omega)$ is
the spin-conserving energy relaxation rate. In that case, electron
transport is much closer to equilibrium in one direction of
current than in another direction of the current, resulting in
asymmetric energy level spectra.~\cite{ralph}
\\\indent It turns out that in spin injection and detection in the regime
defined in this section, $TMR$ is symmetric even if $\Gamma^{L}\ll
\nu^{SF}(\omega)\ll \Gamma^R$. The reason is that the
spin-polarized current is mediated by spin-accumulation, which
takes place after a large number of tunnel-in and tunnel-out
steps.
\\\indent Assume that an electron first tunnels in via the low resistance
junction and then an electron tunnels out via the high resistance
junction. In that case, it is clear that in order to observe spin
polarized current, it is necessary that the spin-relaxation time
in the nanoparticle be longer than the tunnel out time.
\\\indent If the bias voltage is reversed, an electron first tunnels in via
the high resistance junction and then an electron tunnels out via
the low resistance junction. In our regime, it is highly
improbable that the same electron tunnels in and out, because the
number of occupied electron states available for discharge is $\gg
1$. To obtain a spin-polarized current in this case, the
spin-accumulation in the nanoparticle is necessary, which takes
place after many tunnel in and tunnel out steps and makes it
necessary that the spin of the nanoparticle be conserved during
the time that the nanoparticle waits before an electron tunnels
in. Overall, the spin-polarized current and $TMR$ are comparable
in magnitude for the two current directions, in agreement with our
measurements.

\section{Calculation of the Spin Probability Distribution in the Nanoparticle}
\label{cspd}

In this section we obtain the probability distribution among
various many-electron states $|\alpha>$ generated by electron
tunneling via the nanoparticle. The many-electron states are the
Slater determinants of varying single-electron states of the
nanoparticle. In further discussion, the many electron states will
be referred to simply as states.

The number of electrons and the total spin can vary among the
states. The time dependence of the occupational probability
$Q_{\alpha}$ is given by the masters
equation,~\cite{averin1,beenakker1}

\begin{equation}
\frac{dQ_{\alpha}}{dt}=\sum_{{\beta}\neq\alpha}\left(\Gamma_{{\beta}\to
{\alpha}}Q_{\beta}-\Gamma_{{\alpha}\to {\beta}}Q_{\alpha} \right),
\label{master0}
\end{equation}
where $\Gamma_{{\alpha}\to {\beta}}$ is the transition rate from
state $|\alpha>$ to state $|\beta>$.

The steady state solutions are obtained from the masters equation
using $\frac{dQ_{\alpha}}{dt}=0$. The masters equations then
mutually relate the occupational probabilities of various states.
In addition, the occupational probabilities are normalized, that
is, the sum of all occupational probabilities is equal to one. The
masters equation in the steady state and the normalization
condition are sufficient to determine the occupational
probabilities. The current through the left barrier is  obtained
as
\begin{equation}
I_L=|e|\sum_{\alpha} \sum_{\beta} \Gamma_{\alpha\to \beta}^L
Q_\alpha,\label{currentmaster}
\end{equation}
where $\Gamma_{\alpha\to \beta}^L$ is the contribution of the left
lead to the transition rate $\Gamma_{\alpha\to \beta}$, taken with
a positive or negative sign depending on weather the transition
gives a positive or negative contribution to the current,
respectively.~\cite{averin1,beenakker1,bonet}

Because the tunnel density of states in the ferromagnets is
spin-dependent, $Q_{\alpha}$ will depend on the relative magnetic
orientations of the leads.  We set the magnetization of the left
ferromagnetic lead to be always up. The magnetization of the right
ferromagnetic lead can be up or down, corresponding to values of
parameter $\sigma$: if $\sigma = 1$ the magnetizations are
parallel and if $\sigma = -1$ the magnetizations are antiparallel.

 The tunnel-densities of states at the Fermi level of
spin up and spin down electrons in the left lead are $N_\uparrow
=N_{av}(1+P)$ and $N_\downarrow =N_{av}(1-P)$, respectively, where
$N_{av}$ is the average tunnel-density of states at the Fermi
level per spin band, and $P$ is the spin-polarization of the
ferromagnet. Similarly, the tunnel densities of states at the
Fermi level in the right lead are $N_\uparrow =N_{av}(1+\sigma P)$
and $N_\downarrow =N_{av}(1-\sigma P)$ for spin-up and spin-down
electrons, respectively.

We assume that electron spin is conserved in the tunnel process
across a single tunnel junction. This assumption is justified by
the fact that samples without nanoparticles have a large and
weakly voltage dependent $TMR$.~\cite{wei} In that case, the
spin-up (down) bands in the leads are tunnel-coupled only to the
discrete levels in the nanoparticle with spin up (down). This is
also valid because the single-electron states in the nanoparticle
have well defined spin, since the spin-orbit coupling in the
nanoparticle is weak, as discussed earlier.

The tunnel rate between the leads and a discrete level $i$ is
proportional to the tunnel-density of states in the leads and can
be written as $\Gamma_i (1+P)$ for spin-up electrons and $\Gamma_i
(1-P)$ for spin-down electrons, where $\Gamma_i$ is referred to
here as the bare tunnel rate.

Next we use the masters equation to obtain the current in the
regime defined in Sec.~\ref{rmv}.  To summarize, at zero bias
voltage the number of electrons on the nanoparticle is even. We
investigate the region of the IV-curve where only one extra
electron can be added to the nanoparticle, within the first step
of the Coulomb-staircase. This is the bias-voltage region where
the saturation of the spin-polarized current with bias voltage is
observed.  We consider highly asymmetric junctions in resistance,
$R_L\gg R_R$. A positive bias voltage is applied on lead $R$
relative to lead $L$, so first an electron tunnels  into the
nanoparticle from the left lead, across the high-resistance
junction, and after that the nanoparticle discharges one electron
into the right lead via the low resistance junction. Finally, the
internal spin-conserving relaxation rates are much larger than
both tunnel rates, $\nu(\omega)\gg\Gamma^R\gg\Gamma^L$.

In this regime, it is convenient to divide the states into four
groups. Group 1 contains the states which do not have an added
electron and which are fully relaxed with respect to
spin-conserving transitions. We describe these states in detail
before discussing other groups.

\begin{figure}
\includegraphics[width=0.9\textwidth]{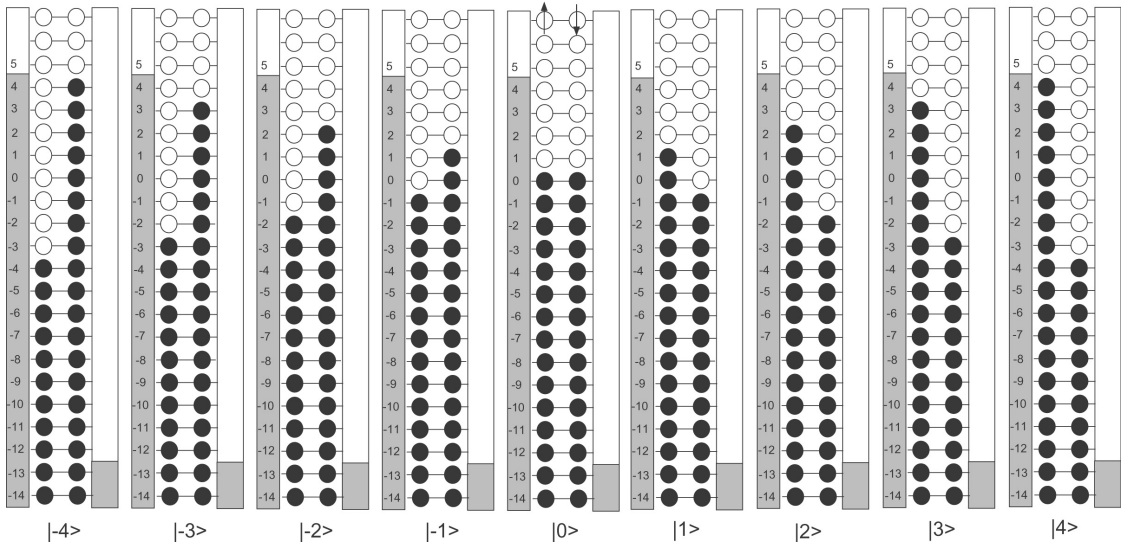}
\caption{Many-electron states from group 1. Black-filled and
white-filled circles indicate occupied and unoccupied
single-electron states, respectively. Only one electron can tunnel
into the nanoparticle at a time because of the Coulomb Blockade;
the displayed states are found before an electron tunnels into the
nanoparticle from the left lead and after an electron tunnelled
out of the nanoparticle into the right lead. The states are fully
relaxed with respect to internal spin-conserving transitions.
~\label{config}}
\end{figure}

Consider the states which do not have an added extra electron
displayed in Fig.~\ref{config}. The internal relaxation of the
states in the figure must involve a spin-flip process. So the
states are fully relaxed with respect to spin-conserving
transitions, thus they are in group 1. In the figure, the charging
energy has been added to the single-electron energy levels of the
nanoparticle, to make it clear which unoccupied single-electron
energy levels are accessible for tunneling in from the left lead.
As seen in the figure, the states from group 1 can be labelled
$|i>$, where $i=-N,-N+1,..., N-1,N$, where $N=4$ for the example
shown in the figure.

$N$ is the number of unoccupied single-electron states above the
Fermi level of the nanoparticle into which an electron can tunnel
in from the left lead. It is related to the bias voltage. If the
level spacing $\delta$ is constant, then
$N=e\frac{C_R}{C_L+C_R}(V-V_{CB})/\delta$,~\cite{wei} where
$C_{L}$ and $C_R$ are the capacitances of the left and the right
tunnel junction, respectively, and $V_{CB}$ is the Coulomb
blockade threshold voltage.~\footnote{If the level spacing
fluctuates, then $N$ is equal to the maximum value of $N'$ for
which $\sum_{i=1}^{N'}\delta_i\leq
e\frac{C_R}{C_L+C_R}(V-V_{CB}).$}

$N$, the number of single-electron states that are accessible for
tunneling-in, is different from the number of single-electron
states that can discharge an electron ($M$). For example, if the
nanoparticle is in state $|0>$, then an electron from the left
lead can tunnel only into one of the unoccupied single-electron
levels labelled $1,2,...,N$  with either spin. After an electron
tunnels in, the nanoparticle can discharge either the added
electron or any other spin-up or spin-down electron from one of
the doubly occupied single-electron states labelled
$0,-1,-2,...,-M$ ($M=12$ in the figure), consistent with Coulomb
blockade.~\cite{averin1,beenakker1} If the level spacing is
constant, then $M=eV_{CB}/\delta$.

Group 2 contains states which do not have an added electron and
which can internally relax by spin-conserving transitions. Group 3
contains states with an added extra electron and which are fully
relaxed with respect to spin-conserving transitions. Finally,
group 4 contains the states with an added extra electron and which
can internally relax by spin-conserving transitions.

The next task is to determine the steady state occupational
probabilities. Using
$\nu(\omega)\gg\Gamma^R\gg\max(\Gamma^L,\nu^{SF}(\omega))$,
several approximations can be made in the masters equation.

First, it is shown in the appendix that the occupational
probabilities $Q_i$ of the states from the first group are larger
than the occupational probabilities of the states from the second,
third, and the forth group, by a factor of
$\frac{\nu(\omega)}{\Gamma^L}$,$\frac{\Gamma^R}{\Gamma^L}$, and
$\frac{\nu(\omega)}{\Gamma^L}$, respectively. The occupational
probabilities of the states from groups 2-4 can thus be neglected
in the normalization condition, so
\[
\sum_{i=-N}^N Q_i = 1,
\]
with an error of order $\frac{\Gamma^L}{\Gamma^R}\ll 1$, in the
limit $\nu(\omega)\gg\Gamma^R\gg\Gamma^L(\omega)$.

Second, if
$\nu(\omega)\gg\Gamma^R\gg\max(\Gamma^L,\nu^{SF}(\omega))$, the
occupational probabilities of the states from the second, third,
and the forth group of states can be eliminated from the masters
equations in an explicit way. The elimination leads to a set of
linear equations that relate the occupational probabilities within
the space of states $G_1$. These equations are refereed to here as
the renormalized masters equation:

\begin{equation}
Q_i \sum_{m\neq i}\Gamma_{i\to m}^{ren}= \sum_{m\neq
i}Q_m\Gamma_{m\to i}^{ren}, \label{master}
\end{equation}
where $\Gamma_{m\to i}^{ren}$ is the renormalized transfer rate
from state $|m>$ to state $|i>$.

The transfer $|m>\to |i>$ takes place either directly, via a
spin-flip transition, or indirectly via intermediate states. In
the leading order of
$\frac{\Gamma^L}{\nu(\omega)}$,$\frac{\Gamma^L}{\Gamma^R}$,
$\Gamma_{m\to i}^{ren}$ is equal to  $\Gamma_{m\to i}$ (direct
transition rate from $m$ to $i$) plus the sum of the rates of
transitions from state $|m>$ into the intermediate states,
weighted by the probability of transfer from the intermediate
states into the state $|i>$:
\begin{equation}
\Gamma_{m\to i}^{ren}=\Gamma_{m\to i}+ \sum_{\alpha}\Gamma_{m\to
\alpha}\Pi(\alpha\to i), \label{gammaeff}
\end{equation}
$\Pi(\alpha\to i)$ is the probability that the nanoparticle in
intermediate state $|\alpha>$ will transfer into state $|i>$.

In the following, we will obtain $\Gamma_{i\to m}^{ren}$ in an
intuitive way. This approach emphasizes understanding and enables
one to obtain the occupational probabilities $Q_i$ from the
renormalized masters equation directly, without explicitly solving
the masters equation. In the appendix, we will derive
$\Gamma_{m\to i}^{ren}$ from the masters equation and show that
the intuitive approach is accurate within controlled
approximations.

\section{Nanoparticle Spin Probability Distribution in Absence of
Spin-Relaxation}

In this section we examine the regime where no spin-relaxation
takes place, $\nu(\omega)\gg\Gamma^R\gg\Gamma^L\gg
\nu^{SF}(\omega)$. In that case, all transfers between the states
within the space of states $|i>$ are indirect. They involve a
sequential electron tunneling process, in which the nanoparticle
goes into an intermediate state with an extra added electron.
$\Gamma_{m\to i}^{ren}$ is equal to the sum of the rates of
transitions from state $|m>$ into the various intermediate states
with an added extra electron, weighted by the probability of
transfer from these intermediate states into the state $|i>$.

Consider the nanoparticle in state $|i>$, which has spin $i\hbar$,
($i>1$). After an electron tunnels into the nanoparticle, the
nanoparticle is in an intermediate state with spin $(i+ 1/2)\hbar$
or $(i- 1/2)\hbar$. Then, after an electron tunnels out, the
nanoparticle spin changes again by $\hbar /2$ or $-\hbar /2$, so
the final spin after a sequential tunneling process can be
 $(i-1)\hbar$, $i\hbar$, and $(i+ 1)\hbar$. So the renormalized transfer rate from
state $|i>$ into states $|m>$ is nonzero only if $m=i-1$ or $m=i$
or $m=i+1$.

The nanoparticle will  transfer  from state $|i>$ to state $|i+1>$
indirectly, if a spin-up electron tunnels in from the left lead
and then a spin-down electron tunnels out into the right lead. A
spin-up electron can tunnel into any of the unoccupied
single-electron states $k$ with spin-up, $k=i+1,i+2, ... N$. After
tunneling in, the nanoparticle instantly relaxes via
spin-conserving transition. A spin-down electron can then tunnel
out from any of the occupied single-electron states $j$ with spin
down, $j=-M, -M+1, ..., -i$. Then, if the nanoparticle after
tunneling out is left in an excited state, it will relax instantly
into state $|i+1>$.

The rate for this spin-up-tunnel-in / spin-down-tunnel-out
sequential process is obtained from Eq.~\ref{gammaeff}, by summing
over all the intermediate states with an added extra electron,
\begin{equation}
\Gamma_{i\to i+1}^{ren}=\sum_{k=i+1}^N\Gamma_k(1+P)
\pi_i(\downarrow) \label{itoi+1}
\end{equation}
where $\pi_i(\downarrow)$ is the probability that the nanoparticle
with an added spin-up electron will discharge a spin-down
electron.

The rate at which a spin-down electron discharges is
$\sum_{j=-M}^{-i}\Gamma_j^R(1-\sigma P)$, where $\Gamma_j^R$ is
the bare tunnel rate between level $j$ and the right lead.
Similarly, the rate at which a spin-up electron discharges is
$\sum_{j=-M}^{i+1}\Gamma_j^R(1+\sigma P)$. The probabilities
$\pi_i(\uparrow)$ and $\pi_i(\downarrow)$ are proportional to the
spin-up and spin-down discharge rates, respectively. The total
discharge probability is one,
$\pi_i(\uparrow)+\pi_i(\downarrow)=1$, so we obtain
\begin{equation}
\pi_i(\downarrow)=\frac{(1-\sigma P)/2}{1+\frac{(1+\sigma
P)\sum_{j=-i+1}^{i+1}\Gamma_j^R}{2\sum_{j=-M}^{-i}\Gamma_j^R}}
\label{pidn}
\end{equation}

As discussed previously, in our model the number of discharging
levels $M\gg 1$. We make a further assumption that $M\gg N$, which
is valid not too far from the  conduction threshold for sequential
electron tunneling through the nanoparticle. In this case, the
denominator in Eq.~\ref{pidn} is close to one, within a factor of
$i/M\sim N/M $, and we obtain

\[
\pi_i(\downarrow)=\frac{1-\sigma
P}{2},\pi_i(\uparrow)=\frac{1+\sigma P}{2}.
\]

The approximation $N/M\ll 1$ enhances the probability to discharge
a spin-down electron at the expense of suppressing the probability
to discharge a spin-up electron. So the approximation increases
the spin accumulation efficiency. This approximation is not
essential for our model to work, but it simplifies further
calculations.

Substituting into Eq.~\ref{itoi+1}, we obtain
\begin{equation}
\Gamma_{i\to i+1}^{ren}=A_{i+1}(P), \label{itoi+2}
\end{equation}
where
\begin{equation}
A_{i}(P)=\frac{(1+P)(1-\sigma P)}{2}\sum_{j=i}^N\Gamma_i .
\label{aip}
\end{equation}

Now we consider the indirect transfer $|i>\to|i-1>$. In this
transfer process, an electron with spin-down tunnels in from the
left lead, followed by a discharge of an electron with spin-up
into the right lead. Following an analysis similar to the above,
we find $\Gamma_{i\to i-1}^{ren}=A_{-i+1}(-P)$. The left hand side
of Eq.~\ref{master} becomes $Q_i(A_{i+1}(P)+A_{-i+1}(-P))$.
 Using
Eqs.~\ref{itoi+2} and ~\ref{itoi-1}, the steady state
Eq.~\ref{master} becomes
\begin{eqnarray}
\lefteqn{Q_i(A_{i+1}(P)+A_{-i+1}(-P))}\nonumber\\
\lefteqn{=} & & Q_{i-1}A_i(P)+Q_{i+1}A_{-i}(-P), \label{master1}
\end{eqnarray} where $i=1,2,...N$.  At $i=N$, in Eq.~\ref{master1}
we must put $A_{N+1}(P)=0$ and $Q_{N+1}=0$.

A similar analysis to the above leads to a steady state equation
for the states with negative nanoparticle spin $|-i>$,
$i=1,2,...,N$:

\begin{eqnarray}
\lefteqn{Q_{-i}(A_{i+1}(-P)+A_{-i+1}(P)}\nonumber\\
\lefteqn{=} & & Q_{-i+1}A_{i}(-P)+Q_{-i-1}A_{-i}(P),
\label{master2} \end{eqnarray} where $i=1,2,...N$. At $i=N$, in
Eq.~\ref{master2} we must put $A_{N+1}(-P)=0$ and $Q_{-N-1}=0$.
The final equation necessary for finding $Q_i$ is
$\sum_{i=-N}^NQ_i=1$.

We calculate the probability distribution numerically for
arbitrary $N$, $\Gamma_i$, and $\Omega_i$, allowing for
fluctuations among different tunnel rates $\Gamma_i$ across the
left tunnel junction. Once the probability distribution is
determined, the current though the nanoparticle is calculated from
Eq.~\ref{currentmaster}, which can be shown to be:

\begin{eqnarray}
\lefteqn{\frac{I}{|e|}=}& &\sum_{\quad i=-N}^NQ_i\frac{I_i}{|e|} = Q_0\sum_{j=1}^N2\Gamma_j+\nonumber\\
& & \quad \sum_{i=1}^NQ_i\left(\sum_{j=i+1}^N2\Gamma_j+\sum_{j=-i+1}^i\Gamma_j(1-P)\right)+\nonumber\\
& & \quad \sum_{i=1}^NQ_{-i}\left(\sum_{j=i+1}^N2\Gamma_j+\sum_{j=-i+1}^i\Gamma_j(1+P)\right)\label{current0}
\end{eqnarray}

\begin{figure}
\includegraphics[width=0.9\textwidth]{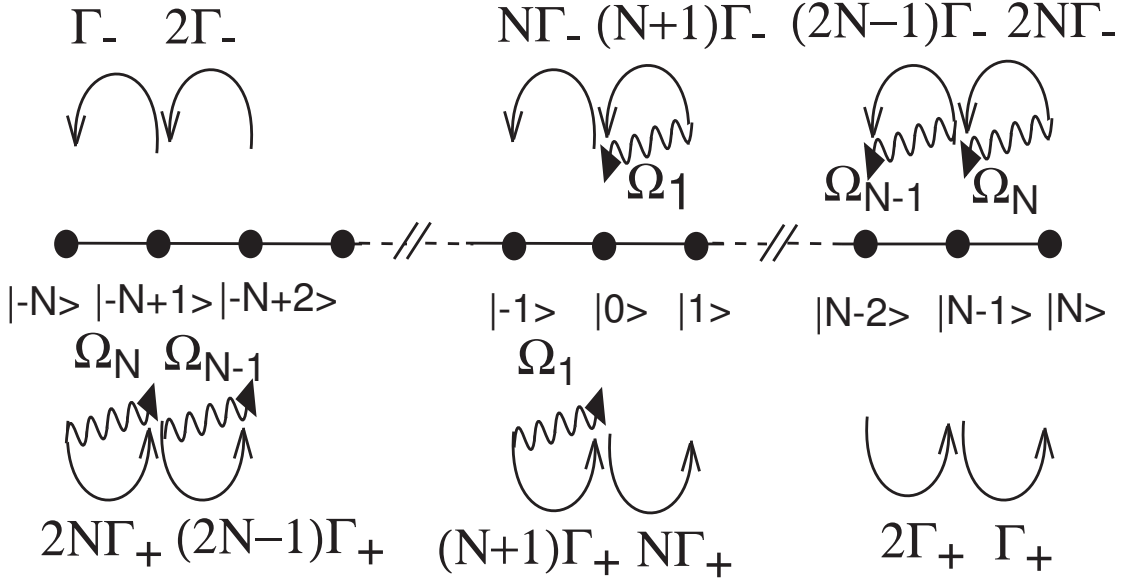}
\caption{Rates of tunnel and internal relaxation transitions
between nanoparticle states $|i>$ in Fig.~\ref{config}, for the
antiparallel magnetic orientations. For simplicity, the bare
tunnel rates to the left lead ($\Gamma_k=\Gamma$)  are assumed to
be independent of single-electron energy levels $k$.
$\Gamma_+=\Gamma (1+P)^2/2$, and $\Gamma_-=\Gamma (1-P)^2/2$. In
the parallel magnetization configuration, substitute  $\Gamma_+$
and $\Gamma_-$ with $\Gamma'=\Gamma (1-P^2)/2$. The wavy line
indicates spin-relaxation processes.~\label{sense}}
\end{figure}

To illustrate these equations, we plot the nanoparticle states
along an axes, as shown in figure~\ref{sense}. We indicate various
transition rates obtained from Eqs.~\ref{itoi+2} and
~\ref{itoi-1}. For simplicity we assume that all bare tunnel-in
rates are the same ($\Gamma_i=\Gamma$). In that case
\begin{equation}
\Gamma_{i\to i+1}^{ren}=(N-i)\frac{(1+P)(1-\sigma P)}{2}\Gamma,
\label{itoi+3}
\end{equation}
and
\begin{equation}
\Gamma_{i\to i-1}^{ren}=(N+i)\frac{(1-P)(1+\sigma P)}{2}\Gamma.
\label{itoi+4}
\end{equation}

The renormalized master equation is
\begin{eqnarray}
\lefteqn{Q_i\left [(N-i)\Gamma(1+P)(1-\sigma P)/2+(N+i)\Gamma(1-P)(1+\sigma P)/2\right]} \nonumber\\
\lefteqn{=} & & Q_{i-1}(N-i+1)\Gamma(1+P)(1-\sigma P)/2+\nonumber\\
& & Q_{i+1}\left((N+i+1)\Gamma(1-P)(1+\sigma P)/2\right),
\label{master3}
\end{eqnarray}
and the current through the nanoparticle, from Eq.~\ref{current0},
becomes
\begin{equation}
I=|e|\left(2N\Gamma-2<i>P\Gamma\right), \label{current2}
\end{equation}
where $<i>=\sum_{i=-N}^NiQ_i$.

One notices from Fig.~\ref{sense} that  the renormalized rate of
transition $|i>\to|i+1>$ decreases as $N-i$, when $i$ increases
from zero to N. Similarly, the renormalized rate of transition
$|i>\to|i-1>$ increases as $N+i$, when $i$ increases from zero to
N. So, if $i$ is near $N$, the rate of $|i>\to|i-1>$ is much
larger than the rate of $|i>\to|i+1>$ and in the steady state,
$Q_i$ must be a rapidly decreasing function of $i$ when $i$ is
near $N$. Similarly, $Q_{-i}$ must be rapidly decreasing with $i$
when $i$ approaches $N$.
\\\indent In the parallel magnetization orientation ($\sigma = 1$) the
transition rates are symmetric around $i=0$, as explained in
Fig.~\ref{sense}. In that case $Q_i$ has a maximum at $i=0$,
indicating that there is no spin-accumulation in the nanoparticle.
This is a situation similar to spin-accumulation in large systems.
\\\indent In the antiparallel configuration of the leads, the rate of
transition $|i>\to|i+1>$ is proportional to $(1+P)^2$ and the the
rate of transition $|i>\to|i-1>$ is proportional to $(1-P)^2$. In
that case, transition rate $|i>\to |i+1>$ is larger than
transition rate $|i+1>\to |i>$ when $i$ is close to zero, leading
to spin-accumulation in the steady state. The spin accumulation is
limited because the rate of transition $|i>\to|i+1>$ decreases in
proportion with $N-i$ and the rate of transition $|i>\to|i-1>$
increases in proportion with $N+i$, as discussed above.

In the following two sections we focus on limit  $N\gg 1$, where
we find the analytical solution of $Q_i$ and $I$. In the analytic
solution we assume that the tunnel rates $\Gamma_i$ across the
left tunnel junction are independent of $i$, $\Gamma_i=\Gamma$.

\subsection{Spin-polarized current at large bias, $N\gg 1$}

At the maximum of $Q_i$ (the mode), which is found at $i=i_0$, the
following condition is satisfied: $\Gamma_{i_0+1\to
i_0}^{ren}=\Gamma_{i_0-1\to i_0}^{ren}$. This leads to
\begin{equation}
i_0=\frac{2PN}{1+P^2}, \label{mode}
\end{equation}
in leading order of $N$ in the antiparallel magnetic configuration
($\sigma=-1$). Thus, in the most probable state of the
nanoparticle, the chemical potential of spin-up electrons is
shifted up by $2i_0\delta$ relative to the chemical potential of
spin-down electrons,
\begin{equation}
\Delta\mu = \frac{4PN\delta}{1+P^2}. \label{deltamu}
\end{equation}

Next we obtain the fluctuations around the mode. We make a
conjecture that the probability distribution $Q_i$ is sharply
peaked around the maximum at $i=i_0$, so that the fluctuations
around $i_0$ are weak compared to $N$. Our numerical calculations
show that $rms(i)\sim\sqrt{N}$, confirming the conjecture. We can
write $i=i_0+j$ and expand $Q_i$ around $i=i_0$:
$Q_i=Q+Q'j+Q''j^2/2+...$ . Substituting into Eq.~\ref{master3}, we
obtain a differential equation, in leading order of $\sqrt{N}$,
\begin{equation}
\frac{N}{2}\left(\frac{1-P^2}{1+P^2}\right)^2\frac{d^2Q}{dj^2}+j\frac{dQ}{dj}+Q=0.
\end{equation}

This linear differential equation can be solved analytically.
Using a boundary condition $Q(\pm N)\ll 1$, the normalized
solution is
\begin{equation}
Q_i\approx\frac{1+P^2}{(1-P^2)\sqrt{\pi
N}}\exp{\left(-\frac{(i-i_0)^2}{N\left(\frac{1-P^2}{1+P^2}\right)^2}\right)},
\label{gaussian}
\end{equation}
which is a Gaussian distribution with fluctuation
\[
rms(i)=\sqrt{\frac{N}{2}}\frac{1-P^2}{1+P^2}.
\]

A similar analysis leads to the probability distribution in the
parallel magnetic configuration:

\begin{equation}
Q_i=\frac{1}{\sqrt{\pi N}}\exp{\left(-\frac{i^2}{N}\right)},
\end{equation}
and $I_{\uparrow\uparrow}=|e|2N\Gamma$, from Eq.~\ref{current2}.

\begin{figure}
\includegraphics[width=0.9\textwidth]{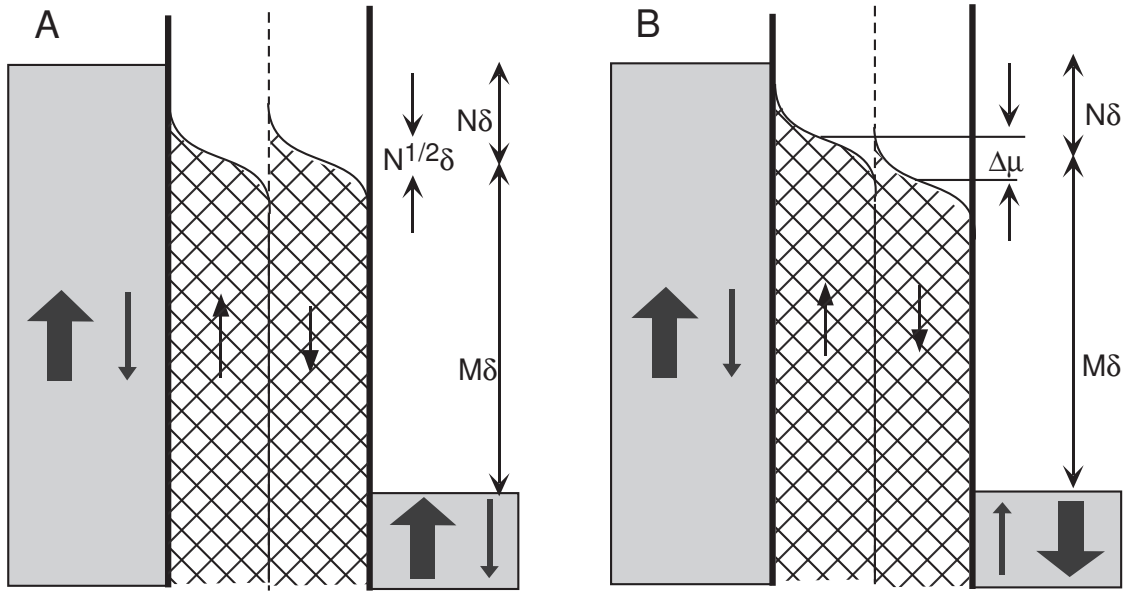}
\caption{A and B: Probability distribution of electrons in
parallel and antiparallel magnetization configuration, when the
number of levels $N$ and $M$ are large.~\label{chemical}}
\end{figure}

Fig.~\ref{chemical} displays the electron distribution function in
the nanoparticle in the parallel and antiparallel magnetization
configurations. The difference in chemical potentials of spin-up
and spin-down electrons in the antiparallel state is proportional
to $N$, the number  of energy levels available for tunneling-in,
according to Eq.~\ref{deltamu}.

Spin accumulation in the nanoparticle is well defined if the
relative fluctuation,
\[
rms(i)/i_0=\frac{1-P^2}{2P\sqrt{2N}},
\]
is smaller than 1. For example, if $P=0.1$, $N$  must be $>12$ in
order to have a well defined spin accumulation. If $N<12$, the
time dependence of the nanoparticle spin will exhibit significant
noise. In large systems where the level spacing is negligibly
small, $N$ is typically $\gg 1$ and the fluctuations are thus
negligible.

Using Eq.~\ref{current2} and the results of this section, the
tunnel magnetoresistance can be shown to be
\begin{equation}
TMR=\frac{P<i>}{N}=\frac{2P^2}{1+P^2}, \label{TMR1}
\end{equation}
which is the Jullieres's formula.~\cite{julliere} The Julliere's
value is larger than the $TMR$ predicted
theoretically.~\cite{weymann2} But this theory is valid in a
different regime than ours and our approximation slightly enhances
the spin-accumulation efficiency, as discussed earlier.  Another
recent numerical calculation~\cite{hwang} in a regime similar to
our own, and an assumption of infinite spin lifetimes confirms the
Julliere value as a maximum of the $TMR$ in a nanoparticle.

\section{Nanoparticle Spin Probability Distribution with
Spin-Relaxation} \label{spinflip}

If spin-relaxation rate is not negligible, then the nanoparticle
in state $|i>$ can undergo a spin-flip transition into state
$|i-1>$. The nanoparticle spin changes by $\hbar$ in a spin-flip
transition, hence there are no other final states in the space of
states $G_1$ after a spin-flip transition. The renormalized
transfer rate is
\begin{equation}
\Gamma_{i\to i-1}^{ren}=A_{-i+1}(-P)+\Omega_i,
 \label{itoi-1}
\end{equation}
where $\Omega_i$ is the total transfer rate from $|i>$ to $|i-1>$
that includes an internal spin flip transition. This transfer can
take place directly or via intermediate states. Starting from a
state $|i>$ in Fig.~\ref{config}, an electron occupying the
highest single-electron energy level $i$ with spin up can make a
transition into the lowest unoccupied single-electron energy level
$-i+1$ with spin down. In this process the energy difference is
$(2i-1)\delta$ and the rate is $\nu^{SF}[(2i-1)\delta]$. This is a
direct transition.

Alternatively, starting from the same sate $|i>$, an electron
occupying the highest single-electron energy level $i$ with
spin-up can make a transition into the next-to-the lowest
unoccupied single-electron energy level $-i+2$ with spin-down. The
nanoparticle is left in an excited state, which is an intermediate
state. This is followed by a spin-conserving relaxation
transition, which is instantaneous in our model, and the
nanoparticle ends in the state $|i-1>$. Similarly, starting from
state $|i>$, an electron occupying the next-to-the highest
single-electron energy level $i-1$ with spin-up can make a
transition into the lowest unoccupied single-electron energy level
$-i+1$ with spin-down, leaving a hole. This is followed again by
an instantaneous spin-conserving relaxation transition, which
brings the nanoparticle into the state $|i-1>$. The energy
differences for the spin-flip processes are the same,
$(2i-2)\delta$, assuming equal level spacing. The total spin-flip
transition rate with this energy difference is
$2\nu_{SF}[(2i-2)\delta]$.

Taking into account all spin-flip processes with varying energy
differences, we find the spin-relaxation rate

\begin{equation}
\Omega_i=\sum_{j=0}^{2i-1}(j+1)\nu_{SF}[(2i-j-1)\delta].
\label{spinfliprate}
\end{equation}

In general, we expect $\nu^{SF}(\omega)$ to be rapidly increasing
with $\omega$: $\nu^{SF}(\omega)\sim\omega^n$. In that case
$\Omega_i$ increases with energy faster than $\nu^{SF}(\omega)$:
$\Omega_i\sim i^{n+2}$.

In this analysis we neglect higher order spin-flip transitions
$|i>\to|i-2>$, $|i>\to|i-3>$,... , where the environment would
receive angular momentum $2\hbar$, $3\hbar$,... , respectively.
For the low energy states of the nanoparticle, we expect that the
probability of the higher order processes to be much smaller than
the probability of the first order processes.

If we assume that spin-relaxation process is governed by phonon
emission and Elliot-Yafet scaling, then
$\Omega_1=\nu^{SF}(\delta)=\alpha\nu_{e-ph}(\delta)$. In that
case,  the spin-relaxation rate increases very rapidly with the
excitation energy. From Eq.~\ref{spinfliprate}, we find
$\Omega_2=46\Omega_1$, $\Omega_3=371\Omega_1$,
$\Omega_4=1596\Omega_1$, and
\begin{equation}
\Omega_i\approx 1.6\alpha\nu_{e-ph}(\delta)i^5=1.6\Omega_1 i^5.
\label{spinfliprate1}
\end{equation}

In the following we will show that the rapid increase of
$\Omega_i$ with $i$ causes the saturation behavior of the
spin-polarized current. It is coincidental that the
spin-relaxation rate $\Omega_i$ of the nanoparticle increases with
fifth-power of the excitation energy, analogous to its temperature
dependence in bulk.

Spin-relaxation in the nanoparticle reduces the spin-accumulation.
Consider again figure~\ref{sense}. The difference between rates
$\Gamma_+$ and $\Gamma_-$ moves the distribution mode from zero to
positive $i$, as discussed in the previous section. The spin
relaxation rate $\Omega_i$ is  added to $\Gamma_-$, reducing the
difference between the upward and downward rates. So the
distribution mode moves downward relative to the mode at
$\Omega_i=0$.

The distribution mode $i_0$ is obtained from $\Gamma_{i_0+1\to
i_0}^{ren}=\Gamma_{i_0-1\to i_0}^{ren}$. It follows that in the
antiparallel state, in leading order of $N$,  the mode satisfies
the following equation
\begin{equation}
N\frac{2P}{1+P^2}=i_0+\frac{\Omega_{i_0}}{\Gamma(1+P^2)}.
\label{transcendent}
\end{equation}

 We set the spin-relaxation rate
$\Omega_1$ to be much smaller than the tunneling rate $\Gamma$,
$\Omega_1\ll \Gamma$, because if $\Omega_1\gg\Gamma$, the
spin-polarized current would be close to zero, contrary to our
measurements. In that case $i_0=N[2P/(1+P^2)]$ for small $N$.

As N increases, both $i_0$ and the spin-relaxation term increase.
The spin-relaxation rate $\Omega_{i}$ increases with $i$ much more
rapidly than $i$ (as $i^{n+2}$). The spin-relaxation begins to
reduce the spin-accumulation when the two terms become comparable,
around
\[
i_0\approx\frac{\Omega_{i_0}}{\Gamma(1+P^2)}\approx
N\frac{P}{1+P^2}.
\]
As $N$ increases further, the spin-relaxation term becomes
dominant, and the mode is obtained from
$\Omega_{i_0}=N[2P/(1+P^2)]$. Assuming $\Omega_i\sim i^{n+2}$ it
follows that at large $N$, which corresponds to large bias
voltage, $i_0\sim N^{1/(n+2)}$. This is a much weaker dependence
on $N$ than $N[2P/(1+P^2)]$. So there is a crossover in
spin-accumulation versus N, from linear dependence to a much
weaker dependence. Similarly, the spin-accumulation crosses over
from liner V-dependence into a much weaker V-dependence at large
V, because N, the number of single-electron energy levels
available for tunneling in, is linear with bias voltage $V$, as
discussed before.

One consequence of the rapid increase of $\Omega_i$ with $i$ is an
asymmetric spin probability distribution. Our numerical
calculations show that the probability that the nanoparticle spin
is below the mode is larger than the probability that the
nanoparticle spin is above the mode. In that case, the average
nanoparticle spin is smaller than the most probable nanoparticle
spin, $<i> <i_0$. Nevertheless, as long as the width of the
distribution is much smaller than the mode, one can substitute
$i_0$ for $<i>$ in Eq.~\ref{current2} and the spin polarized
current
$I_{\uparrow\uparrow}-I_{\uparrow\downarrow}=2|e|P<i>\Gamma\approx
2|e|Pi_0$ exhibits the same crossover with bias voltage. This
explains the saturation of spin-polarized current with bias
voltage observed in our experiments.

The crossover condition can be rewritten as
$\Omega_{i_0}=NP\Gamma$. So, at the crossover point, the rate of
the spin-flip transition with energy difference $\Delta\mu$
(Eq.~\ref{deltamu}) corresponds to the spin-polarized current,
e.g. $\Omega(\Delta\mu)\approx
(I_{\uparrow\uparrow}-I_{\uparrow\downarrow})/|e|$.

\begin{figure}
\includegraphics[width=0.9\textwidth]{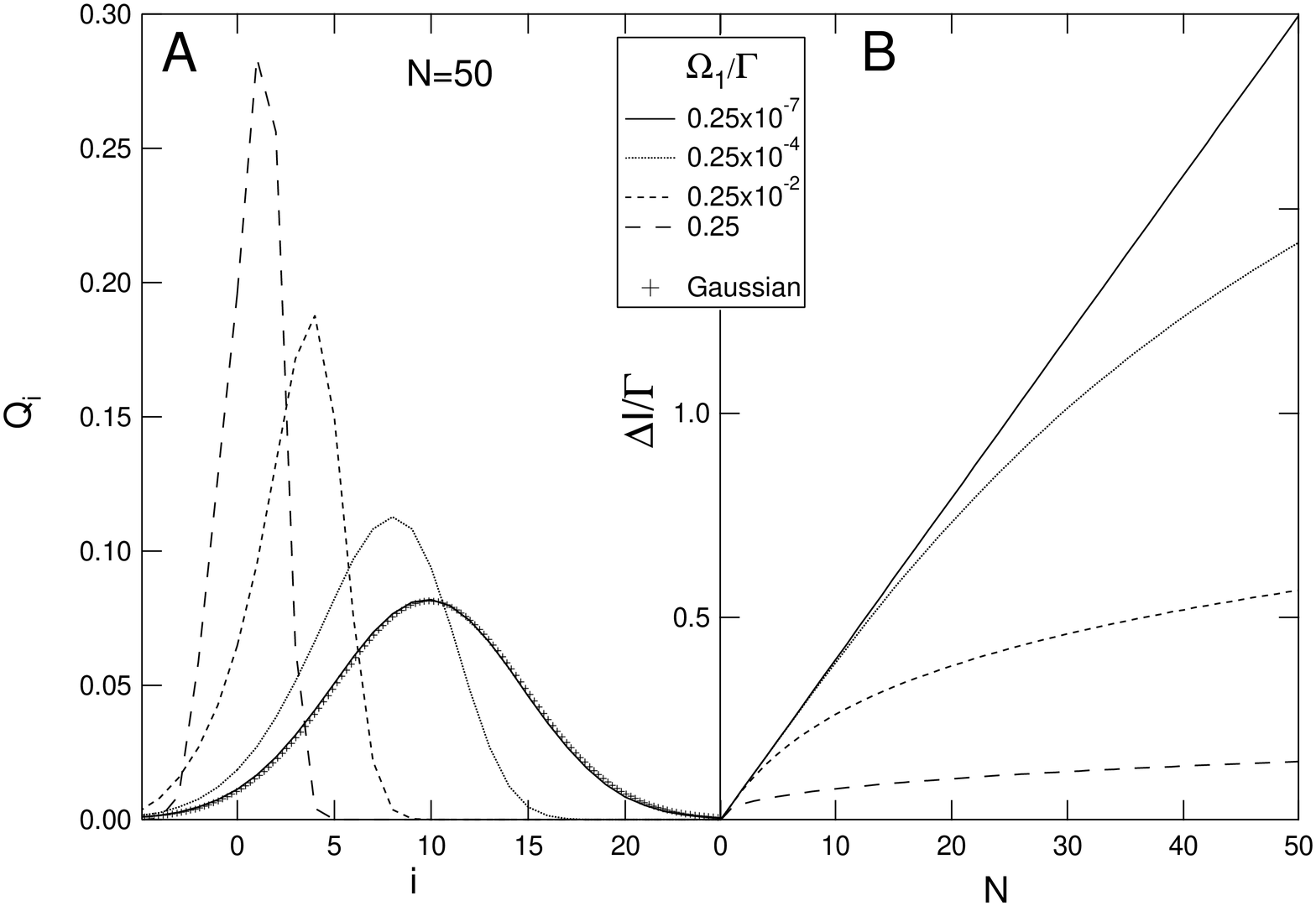}
\caption{A: Probability distribution functions of the nanoparticle
many-electron states in the antiparallel state, when the number of
energy levels available to tunnel in is $N=50$. B: Spin-polarized
current $\Delta I=I_{\uparrow\uparrow}-I_{\uparrow\downarrow}$
versus $N$, the number of energy levels available to tunnel in.
$N$ and bias voltage are related linearly.~\label{distro}}
\end{figure}

Now we discuss the spin-relaxation effects assuming that the
relaxation is mediated by phonon emission, in accordance with the
Elliot-Yafet relation Eq.~\ref{spinfliprate1}.
Figure~\ref{distro}-A displays the distribution function at large
bias voltage, $N=50$, obtained by numerical calculations for
$P=0.1$. $\Omega_1/\Gamma$ is varied from $0.25$ to $0.25\cdot
10^{-7}$. Also shown is the Gaussian distribution in
Eq.~\ref{gaussian}, which is valid in absence of any
spin-relaxation. At $\Omega_1=0.25\cdot 10^{-7}$ the probability
distribution is very close to the Gaussian, indicating that the
spin relaxation is negligible. As $\Omega_1/\Gamma$ increases, the
mode shifts downwards and the distribution becomes asymmetric. At
$\Omega_1/\Gamma = 0.25$, the mode is located at $i_0=1$, showing
that there the spin accumulation is very weak.

The distribution mode is now obtained from
\begin{equation}
N\frac{2P}{1+P^2}= i_0+\frac{1.6\Omega_1 i_0^5}{\Gamma(1+P^2)}.
\label{transcendent1}
\end{equation}
At large N, when the spin relaxation term dominates,
\[
i_0 \approx \left( \frac{2PN\Gamma}{1.6\Omega_1}\right)^{1/5},
\]
that is, the mode crosses over from being linear with $N$ to being
proportional to $N^{1/5}$.

Fig.~\ref{distro}-B displays spin-polarized current versus $N$
obtained by numerical calculations, for different
$\Omega_1/\Gamma$. The crossover from linear to a much weaker
dependence is evident for $\Omega_1/\Gamma =0.25$ and $0.25\cdot
10^{-2}$.
The crossovers in $i_0$ and
$I_{\uparrow\uparrow}-I_{\uparrow\downarrow}$ are equivalent, as
discussed before, so at large $N$,
$I_{\uparrow\uparrow}-I_{\uparrow\downarrow}\sim N^{1/5}$.

At $\Omega_1/\Gamma =0.25$, the rate of spin-relaxation with
energy difference $\delta$ is small compared to the tunneling
rate. But $\Omega_2/\Gamma =11.5$, so the spin-relaxation rate
with an energy difference $\geq 2\delta$ is large compared to the
tunneling rate. As a result, the spin-polarized current crosses
over already at the second single electron energy level. In
particular, we find that the contribution from the second
single-electron energy level to the spin polarized current is
approximately a 3rd of the contribution from the first
single-electron energy level.

\section{Fitting}
\label{fit}

\begin{figure}
\includegraphics[width=0.9\textwidth]{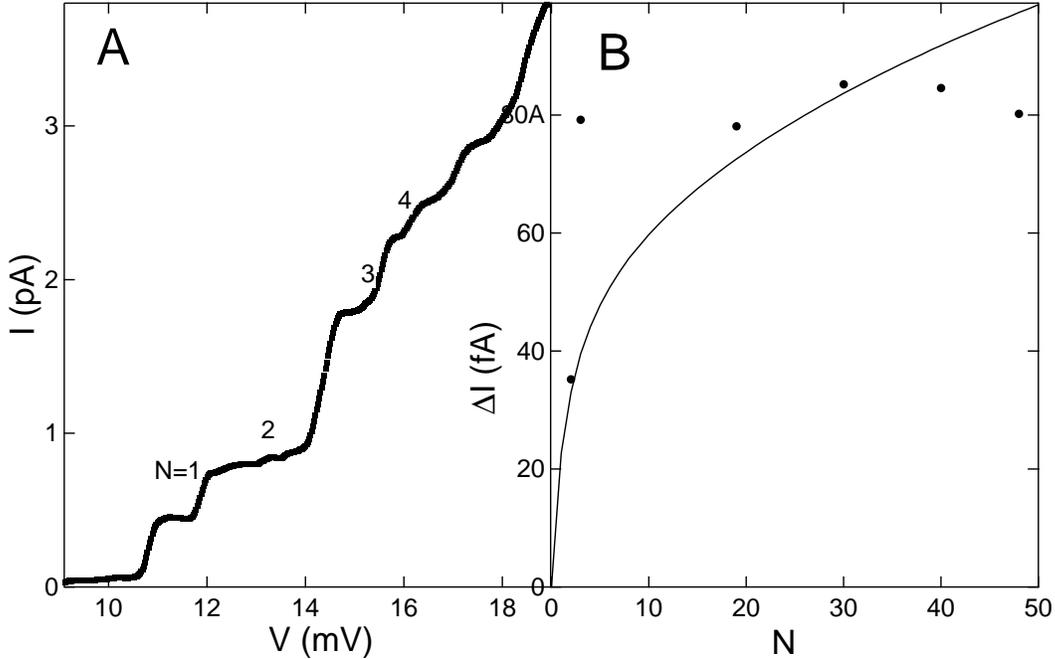}
\caption{A: I-V curve in sample 1 near the sequential tunneling
threshold.  B: Circles: Spin polarized current $\Delta
I=I_{\uparrow\uparrow}-I=I_{\uparrow\downarrow}$ versus $N$. Line
is the best fit to the model.~\label{expdata}}
\end{figure}

To illustrate how the model derived in this paper can be used to
interpret measurements of the spin-polarized current, we discuss
sample 1 in Ref.~\cite{wei}. In Fig.~\ref{expdata}-A we display
the I-V curve at $T=0.03K$, obtained in the parallel magnetization
configuration and in the regime where the electron discharge rate
is much faster than the tunnel in rate, as required by our model
(we have reversed the sign of bias voltage compared to that in
Ref.~\cite{wei}). The I-V curve increases in discrete steps at
voltages corresponding to discrete energy levels of the
nanoparticle. The average tunnel in rate is obtained as $\Gamma
=<\delta I>/2|e|=1.5MHz$,~\cite{wei} where $<\delta I>$ is the
average current step. The average level spacing $\delta=0.8meV$
corresponds to a spherical nanoparticle with diameter 6nm. The
electron g-factors are very close to 2,~\cite{wei} confirming weak
spin-orbit scattering.

The spin polarized current, $\Delta
I=I_{\uparrow\uparrow}-I_{\uparrow\downarrow}$, versus $N$ is
shown by circles in Fig.~\ref{expdata}-B. At small $N$, energy
levels are well resolved, and $N$ is observed directly as shown in
the figure. At large bias voltage, where energy levels are
broadened, we find $N$ as $ec(V-V_{CB})/\delta$,~\cite{wei} where
$c$ is the capacitance ratio that converts from voltage to
nanoparticle energy and $V_{CB}$ is the Coulomb blockade voltage
threshold.

We fit $\Delta I$ obtained from the model versus $N$, using a
fixed tunnel rate $\Gamma=1.5MHz$ and two fit parameters, $P$ and
$\Omega_1$. We assume $\Omega_i\approx 1.6i^5\Omega_1$, which is
valid when spin relaxation is mediated by phonons following
Elliot-Yafet scaling, as discussed earlier. The best fit
parameters are $P=0.23$ and $\Omega_1=1.1\Gamma\approx 1.6MHz$.
This value of the spin-relaxation rate is consistent with what we
estimated based on qualitative discussions in Ref.~\cite{wei}

The best fit does not fully saturate with bias voltage, as seen in
Fig.~\ref{distro}-B and ~\ref{expdata}-B, but exhibits a crossover
to a weaker dependence with $N$. Nevertheless, we find the
agreement between the model and the data to be qualitatively good.
An outlying point at $N=3$ in Fig.~\ref{expdata}-B is attributed
to the large tunnel rate via the third energy levels as seen in
Fig.~\ref{expdata}-A, because we use a constant tunnel in rate in
the fit. Such a large tunnel in rate arises from the natural
statistical fluctuations of the tunnel rates among different
levels.

If we assume that the spin-relaxation rate is energy independent,
$\Omega_i=\Omega_1$, the spin-polarized current in our model will
have a linear dependence with $N$. In that case,  not only does
the best fit have five times larger chi-square, but also the best
fit parameters are un-physical, $P=0.045$ and
$\Omega_1=0.067\Gamma$. In particular, the measured spin-polarized
current via the low energy levels would be 25 times larger than
the maximum theoretical value corresponding to $TMR=2P^2/(1+P^2)$
at $P=0.045$. Consequently, an energy independent spin-relaxation
rate cannot explain our results, and the energy dependence of the
spin-relaxation rate plays a critical role in interpreting the
spin-polarized current through the nanoparticle in the regime of
well resolved energy levels and weak spin-orbit scattering.

Our model neglects higher order spin relaxation processes. We
suggest that there could also be transitions where two electrons
flip their spin simultaneously, emitting a phonon with angular
momentum $2\hbar$. The probability of such a transition path must
be much weaker  relative to the probability of the 1st order
process (weaker by a factor of $\alpha$) . But with increasing
$i$, the number of transition paths behind a second-order process
would increase much faster with $i$ than in $\Omega_i$, leading to
a larger exponent in the dependence of the spin-relaxation rate on
$i$. As a result, at high excitation energies, second order or
even higher order spin-relaxation process can dominate
spin-relaxation, leading to another crossover in $\Delta I$ versus
$V$, weakening the $V$ dependence further. Thus, inclusion of the
the higher order spin-relaxation processes at high excitation
energies would improve the agreement between model and
measurements.

\section{Conclusion}

In conclusion, we have derived a simple model for calculating
spin-accumulation and spin-polarized current via discrete energy
levels of a metallic nanoparticle in the regime of weak spin-orbit
scattering. The model is valid in a relatively narrow range of
sample parameters. However, a large percentage of samples
fabricated by lithography have parameters in that range. The
energy dependence of the spin-relaxation rate causes a significant
suppression of the bias voltage dependence of the spin-polarized
current at large bias voltage. In particular, if the
spin-relaxation rate increases with excitation energy as a power
of $n$, $\nu^{SF}(\omega)\sim\omega^n$, then the spin polarized
current at large bias voltage $V$ will increase proportionally to
$V^{1/(n+2)}$, which is a dependence much weaker than linear. The
crossover between the linear dependence at low voltage and the
much weaker dependence at large voltage occurs when the the
spin-polarized current$/|e|$ is equal to the spin-relaxation rate
with the energy difference given by the spin-accumulation. The
model leads to the spin-relaxation rate $1.6 MHz$ in an Aluminum
nanoparticle of diameter $6nm$, for a transition with an energy
difference of one level spacing.

This work was performed in part at the Georgia-Tech electron
microscopy facility. This research is supported by the DOE grant
DE-FG02-06ER46281 and David and Lucile Packard Foundation grant
2000-13874.

\section{Appendix: Masters Equation}

In the steady state the occupational probabilities are
time-independent and the masters equation~\ref{master0} becomes
\begin{equation}
Q_{\alpha}\sum_{{\beta\neq\alpha}}\Gamma_{{\alpha}\to
{\beta}}=\sum_{{\beta}\neq\alpha}Q_{\beta}\Gamma_{{\beta}\to
{\alpha}}.
 \label{mastera0}
\end{equation}
The occupational probabilities satisfy the normalization condition
$\sum_\alpha Q_{\alpha}=1$.

In this section we reduce the problem of solving
Eq.~\ref{mastera0} into a simpler problem. This is done by
identifying a group of states which have much larger probability
than the states outside the group and then eliminating the
occupational probabilities of the states outside the group.

We use inequalities $\nu_{\alpha\to\beta}\gg\Gamma^R\gg\max
(\Gamma^L,\nu_{\alpha\to\beta}^{SF})$. $\nu_{\alpha\to\beta}$ is
the spin-conserving relaxation rate from state $|\alpha>$ to state
$|\beta>$ given by Eq~\ref{eph}, in which
$\omega=E_{\alpha}-E_{\beta}$ is the energy difference between the
initial and the final state.  We calculate at zero temperature, so
$\nu_{\alpha\to\beta}=0$ if $E_{\alpha}<E_{\beta}$.
$\nu_{\alpha\to\beta}$ is nonzero only if the initial and the
final state have the same spin. In this regime the electron
transport through the nanoparticle is in equilibrium with respect
to spin-conserving relaxation.

 $\nu_{\alpha\to\beta}^{SF}$ is the
spin-flip relaxation rate between the states. It can be calculated
using Eq.~\ref{eliot}. $\nu_{\alpha\to\beta}^{SF}$ is nonzero only
if $S_{\beta}^z-S_{\alpha}^z=\pm\hbar$.

In the analysis, we separate the states that are fully relaxed
with respect to spin-conserving relaxation processes from those
that are not. The latter states decay at rate
$\nu_{\alpha\to\beta}$, whereas the former states decay at a rate
$\max {(\Gamma^L,\nu^{SF})}$ or $\Gamma^R$, depending on weather
the number of electrons is even (before tunneling in after
tunneling out) or odd (after tunneling in before tunneling out).
From Eq.~\ref{mastera0} it follows that the occupational
probabilities of states that are not fully relaxed with respect to
spin-conserving transitions are strongly suppressed in the limit
$\nu_{\alpha\to\beta}\gg\Gamma^R\gg\max
(\Gamma^L,\nu_{\alpha\to\beta}^{SF})$, because the occupational
probability of those states have $\nu_{\alpha\to\beta}$ in the
denominator.

We are considering only the states with one extra added electron,
as discussed in sections ~\ref{rmv} and ~\ref{cspd}. Let $Q_{i}$
denote the occupational probabilities of the states without an
added extra electron that are fully relaxed by spin-conserving
transitions. These states are from group $G_1$ and are shown in
Fig.~\ref{config}. Let $P_j$ denote the occupational probabilities
of the states with an added extra electron that are fully relaxed
by spin-conserving transitions; these states are from group $G_3$.

Next we consider the excited states. $Q'_\alpha$ denotes the
occupational probabilities of the states without an added electron
that are not fully relaxed by spin-conserving transitions; these
are the states from group $G_2$. Finally, $P'_\beta$ denotes the
occupational probabilities of the states with an added extra
electron that are not fully relaxed by spin-conserving
transitions; these are states from group $G_4$.

The masters equations are next written for the states in each
group. For groups 1 and 2, the equations are

\begin{eqnarray}
\nonumber
 &Q_i\bigg [ \sum_{m\in G_1}\nu_{i\to m}^{SF}+\sum_{\alpha'\in
G_2}\nu_{i\to \alpha'}^{SF}&\\
\nonumber &+\sum_{j'\in G_3}\Gamma_{i\to j'}+\sum_{\beta'\in
G_4}\Gamma_{i\to \beta'}\bigg ] &\\
\nonumber &= \sum_{m\in G_1}Q_m\nu_{m\to i}^{SF}+\sum_{\alpha'\in
G_2}Q'_{\alpha'}(\nu_{\alpha'\to i}+\nu_{\alpha'\to i}^{SF})  &\\
&+\sum_{j'\in G_3}P_{j'}\Gamma_{j'\to i}+\sum_{\beta'\in
G_4}P'_{\beta'}\Gamma_{\beta'\to i},& \label{msa1}
\end{eqnarray}

\begin{eqnarray}
\nonumber
 &Q'_\alpha\bigg [ \sum_{m\in G_1}(\nu_{\alpha\to m}+\nu_{\alpha\to m}^{SF})+\sum_{\alpha'\in
G_2}(\nu_{\alpha\to \alpha'}+\nu_{\alpha\to \alpha'}^{SF})&\\
\nonumber &+\sum_{j'\in G_3}\Gamma_{\alpha\to j'}+\sum_{\beta'\in
G_4}\Gamma_{\alpha\to \beta'}\bigg ] &\\
\nonumber &= \sum_{m\in G_1}Q_m\nu_{m\to
\alpha}^{SF}+\sum_{\alpha'\in
G_2}Q'_{\alpha'}(\nu_{\alpha'\to \alpha}+\nu_{\alpha'\to \alpha}^{SF})&\\
&+\sum_{j'\in G_3}P_{j'}\Gamma_{j'\to \alpha}+\sum_{\beta'\in
G_4}P'_{\beta'}\Gamma_{\beta'\to \alpha},& \label{msa2}
\end{eqnarray}
where $i\in G_1$, $\alpha\in G_2$. Rates $\Gamma_{i\to j'}$,
$\Gamma_{i\to \beta'}$, $\Gamma_{\alpha\to j'}$, and
$\Gamma_{\alpha\to \beta'}$ on the left hand sides are
proportional to the tunnel-in rate $\Gamma^L$ across the left
junction. Similarly, rates $\Gamma_{j'\to i}$, $\Gamma_{\beta' \to
i}$, $\Gamma_{\alpha\to j'}$, and $\Gamma_{\beta' \to \alpha}$ on
the right hand sides are proportional to the tunnel-out rate
$\Gamma^R$ across the left junction.

For groups 3 and 4, the equations are

\begin{eqnarray}
\nonumber
 &P_j\bigg [ \sum_{m\in G_1}\Gamma_{j \to m}+\sum_{\alpha'\in
G_2}\Gamma_{j \to \alpha'}&\\
\nonumber &+\sum_{j'\in G_3}\nu_{j\to j'}^{SF}+\sum_{\beta'\in
G_4}\nu_{j\to \beta'}^{SF}\bigg ] &\\
\nonumber &= \sum_{m\in G_1}Q_m\Gamma_{m\to j}+\sum_{\alpha'\in
G_2}Q'_{\alpha'}\Gamma_{\alpha' \to j}+  &\\
&\sum_{j'\in G_3}P_{j'}\nu_{\j'\to j}^{SF}+\sum_{\beta'\in
G_4}P'_{\beta'}(\nu_{\beta'\to j}+\nu_{\beta'\to j}^{SF}),&
\label{msa3}
\end{eqnarray}
\begin{eqnarray}
\nonumber
 &P'_{\beta}\bigg [ \sum_{m\in G_1}\Gamma_{\beta \to m}+\sum_{\alpha'\in
G_2}\Gamma_{\beta \to \alpha'}&\\
\nonumber &+\sum_{j'\in G_3}(\nu_{\beta\to j'}+\nu_{\beta\to
j'}^{SF})+\sum_{\beta'\in
G_4}(\nu_{\beta\to \beta'}+\nu_{\beta\to \beta'}^{SF})\bigg ] &\\
\nonumber &= \sum_{m\in G_1}Q_m\Gamma_{m\to
\beta}+\sum_{\alpha'\in
G_2}Q'_{\alpha'}\Gamma_{\alpha' \to \beta}+  &\\
&\sum_{j'\in G_3}P_{j'}\nu_{\j'\to \beta}^{SF}+\sum_{\beta'\in
G_4}P'_{\beta'}(\nu_{\beta'\to \beta}+\nu_{\beta'\to
\beta}^{SF}),& \label{msa4}
\end{eqnarray}
where $j\in G_3$, $\beta\in G_4$.

First we estimate the orders of magnitude of the occupational
probabilities. The transition rates are assumed roughly constant
to within an order of magnitude and pulled out of the sum, and
$Q$, $Q'$, $P$, and $P'$ refer to the probability of finding the
particle in a state from group 1, 2,  3, and 4,
respectively.~\footnote{The internal transition probabilities
depend on the energy of the states. It can be shown that our
estimations of the occupational probabilities remain valid if the
energy dependence is included, provided that
$\nu_{\alpha\to\beta}\gg\Gamma^R\gg\max
(\Gamma^L,\nu_{\alpha\to\beta}^{SF})$.} Entering the orders of
magnitude of various terms in Eq.~\ref{msa2}, we obtain
$Q'(\nu+\nu^{SF}+\Gamma^L)\sim
Q\nu^{SF}+Q'(\nu+\nu^{SF})+P\Gamma^R+P'\Gamma^R$, where $\nu$ and
$\nu^{SF}$ indicate orders of magnitude of $\nu_{\alpha\to\beta}$
and $\nu_{\alpha\to\beta}^{SF}$, respectively. In the limit
$\nu\gg\Gamma^R\gg\max (\Gamma^L,\nu^{SF})$, one gets $Q'\nu\sim
Q\nu^{SF} +(P+P')\Gamma^R$.

Similarly, substituting orders of magnitude in Eq.~\ref{msa1}, one
obtains $Q(\nu^{SF}+\Gamma^L)\sim
Q\nu^{SF}+Q'(\nu+\nu^{SF})+P\Gamma^R+P'\Gamma^R$. In the limit
$\nu\gg\Gamma^R\gg\max (\Gamma^L,\nu^{SF})$, one gets
$Q(\nu^{SF}+\Gamma^L)\sim Q'\nu+(P+P')\Gamma^R$. Combining the two
order of magnitude estimates we obtain
\begin{equation}
Q'\sim Q\frac{\nu^{SF}+\Gamma^L}{\nu}\approx
Q\frac{\max(\nu^{SF},\Gamma^L)}{\nu}\ll Q.\label{est1}
\end{equation}
Hence, the occupational probabilities $Q_\alpha'$  of the excited
states from group $G_2$ are much smaller than the occupational
probabilities $Q_i$ of states that are fully relaxed with respect
to spin-conserving transitions, an expected result.

Next, we substitute into Eq.~\ref{msa4} the orders of magnitude of
various terms and obtain $P'(\Gamma^R+\nu+\nu^{SF})\sim
Q\Gamma^L+Q'\Gamma^L+P\nu^{SF}+P'(\nu+\nu^{SF})$. Using
$\nu\gg\Gamma^R\gg\max (\Gamma^L,\nu^{SF})$ and Eq.~\ref{est1},
one obtains $P'\nu\sim Q\Gamma^L+P\nu^{SF}$. Finally,
Eq.~\ref{msa3} leads to order of magnitude estimate
$P(\Gamma^R+\nu+\nu^{SF})\sim
Q\Gamma^L+Q'\Gamma^L+P\nu^{SF}+P'(\nu+\nu^{SF})$, which leads to
$P\Gamma^R\sim Q\Gamma^L+P'\nu$. Combining the estimates, we
obtain
\begin{equation}
P\sim Q\frac{\Gamma^L}{\Gamma^R}\ll Q, P'\sim P\frac{
\Gamma^R}{\nu}\ll P. \label{est2}
\end{equation}

So, the occupational probabilities of states with an added
electron and fully relaxed with respect to spin-conserving
transitions are much smaller than the occupational probabilities
of states without an added electron and fully relaxed with respect
to spin-conserving transitions, as expected from the large
asymmetry in tunnel resistance. In addition, the occupational
probabilities of excited states with an added electron are much
smaller than the occupational probabilities of states with an
added electron and fully relaxed with respect to spin-conserving
transitions.

Using the estimates in Eq.~\ref{est1} and ~\ref{est2}, the leading
order of the masters equation for groups 1 and 2 are

\begin{eqnarray}
\nonumber
 &Q_i\bigg [ \sum_{m\in G_1}\nu_{i\to m}^{SF}+\sum_{\alpha'\in
G_2}\nu_{i\to \alpha'}^{SF}&\\
\nonumber &+\sum_{j'\in G_3}\Gamma_{i\to j'}+\sum_{\beta'\in
G_4}\Gamma_{i\to \beta'}\bigg ] &\\
\nonumber &= \sum_{m\in G_1}Q_m\nu_{m\to i}^{SF}+\sum_{\alpha'\in
G_2}Q'_{\alpha'}\nu_{\alpha'\to i}  &\\
&+\sum_{j'\in G_3}P_{j'}\Gamma_{j'\to i},& \label{msb1}
\end{eqnarray}

\begin{eqnarray}
\nonumber
 &Q'_\alpha\bigg [ \sum_{m\in G_1}\nu_{\alpha\to m}+\sum_{\alpha'\in
G_2}\nu_{\alpha\to \alpha'}\bigg ]&\\
\nonumber &= \sum_{m\in G_1}Q_m\nu_{m\to
\alpha}^{SF}+\sum_{\alpha'\in
G_2}Q'_{\alpha'}\nu_{\alpha'\to \alpha}&\\
&+\sum_{j'\in G_3}P_{j'}\Gamma_{j'\to \alpha},& \label{msb2}
\end{eqnarray}
where $i\in G_1$, $\alpha\in G_2$. For groups 3 and 4, the
equations  are

\begin{eqnarray}
\nonumber
 &P_j\bigg [ \sum_{m\in G_1}\Gamma_{j \to m}+\sum_{\alpha'\in
G_2}\Gamma_{j \to \alpha'} \bigg ]&\\
 &= \sum_{m\in G_1}Q_m\Gamma_{m\to j}+\sum_{\beta'\in
G_4}P'_{\beta'}\nu_{\beta'\to j},& \label{msb3}
\end{eqnarray}

\begin{eqnarray}
\nonumber
 &P'_{\beta}\bigg [ \sum_{j'\in G_3}\nu_{\beta\to j'}+\sum_{\beta'\in
G_4}\nu_{\beta\to \beta'}\bigg ] &\\
 &= \sum_{m\in G_1}Q_m\Gamma_{m\to \beta}
+\sum_{\beta'\in G_4}P'_{\beta'}\nu_{\beta'\to \beta},&
\label{msb4}
\end{eqnarray}
where $j\in G_3$, $\beta\in G_4$. The relative errors of the terms
in these equations are smaller than the maximum of
$\max(\Gamma^L,\nu^{SF})/\nu$, $\Gamma_L/\Gamma_R$,
$\nu^{SF}/\Gamma_R$, and $\Gamma_R/\nu$.

Now we begin the process of elimination. The first step is to
eliminate the excited states. In principle, the space of excited
states is very large, if the excited states include multiple
electron hole pairs.~\cite{agam} These multiply excited states are
generated by tunneling if the particle remains excited longer than
the time of a sequential tunneling cycle, so in our limit the
occupational probabilities of the excited states with multiple
electron-hole pairs are negligibly small. In this regime, the
space of excited states $G_2$ can be restricted to the excited
states that can undergo a direct spin-conserving transition into a
state from $G_1$. Similarly, the space of excited states $G_4$ can
be restricted to those excited states that can undergo a direct
spin-conserving transition into a state from $G_3$.

Consider Eq.~\ref{msb4}, which represents an equilibrium condition
for a state $|\beta>$, $\beta\in G_4$. For a given state $|j>$,
$j\in G_3$, we select a subspace within $G_4$, such that the
nanoparticle in a state from the subspace can relax via
spin-conserving transition into the state $|j>$. That is, for any
state $|\beta>$ within the subspace, $\nu_{\beta\to j}\neq 0$, and
for any state $|\beta>$ outside the subspace, $\nu_{\beta\to j}=
0$. Then we sum Eq.~\ref{msb4} over the subspace, which leads to
\begin{eqnarray}
\nonumber
 &\sum_{\beta\in G_4}'\sum_{j'\in G_3}P'_{\beta}\nu_{\beta\to j'}+\sum_{\beta\in G_4}'\sum_{\beta'\in
G_4}P'_{\beta}\nu_{\beta\to \beta'}=&\\
\nonumber & \sum_{m\in G_1}\sum_{\beta\in G_4}'Q_m\Gamma_{m\to
\beta} +\sum_{\beta\in G_4}'\sum_{\beta'\in
G_4}P'_{\beta'}\nu_{\beta'\to \beta},&
\end{eqnarray}
where $\sum_{\beta\in G_4}'$ sums only over the states within the
subspace, so that $\nu_{\beta\to j}\neq 0$.

In the second term on the left hand side of this equation, where
$\beta$ is restricted within the subspace, $\beta'$ becomes also
restricted within the subspace, because of the spin-conservation
(since states $|\beta>$ and $|j>$ have the same spin, and states
$|\beta'>$ and $|\beta>$ also have the same spin, $\beta'$ must be
restricted within the subspace). The second term becomes
 $\sum_{\beta,\beta'\in
G_4}'P'_\beta\nu_{\beta\to\beta'}$. Similarly, in the second term
on the right hand side, $\nu_{\beta'\to\beta}$ is nonzero only if
states $|\beta'>$ and $|\beta>$ have the same spin as the spin in
state $|j>$, and the term becomes $\sum_{\beta,\beta'\in
G_4}'P'_{\beta'}\nu_{\beta'\to\beta}$. Exchanging the indices
$\beta$ and $\beta'$, the second terms on the left hand side and
the right hand side cancel.

Now consider the first term on the left hand side. As stated
above, only those states $|\beta>$ that can undergo a
spin-conserving transition into the state $|j>$ are in the sum
over $\beta$. It follows that $\nu_{\beta\to j'}$ is nonzero only
if $j'=j$, since varying states $|j'>$ from space $G_3$ have
different spin. The first term on the left hand side becomes
$\sum_{\beta\in G_4}'P'_\beta \nu_{\beta\to j}$. We can remove the
prime and sum instead over the entire space $G_4$, because
$\nu_{\beta\to j}$ automatically restricts the sum to the
subspace. One obtains

$ \sum_{\beta\in G_4}P'_\beta \nu_{\beta\to j}=\sum_{m\in G_1}Q_m
\sum_{\beta\in G_4}'\Gamma_{m\to \beta}. $

The  left hand side in this equation is the same as the second
term on the right hand side of Eq.~\ref{msb3} and thus it can  be
eliminated, which leads to

\begin{eqnarray}
P_j=\sum_{m\in G_1}Q_m\frac{\Gamma_{m\to j}+\sum_{\beta\in
G_4}'\Gamma_{m\to \beta}}{\sum_{m\in G_1}\Gamma_{j \to
m}+\sum_{\alpha'\in G_2}\Gamma_{j \to \alpha'}} .\label{PvsQ}
\end{eqnarray}

Next we perform a similar elimination process using
Eqs.~\ref{msb1} and ~\ref{msb2}. For a given $i\in G_1$, we sum
Eq.~\ref{msb2} over the excited states $|\alpha>$, $\alpha\in
G_2$, for which $\nu_{\alpha\to i}\neq 0$,
\begin{eqnarray}
\nonumber
 &\sum_{\alpha\in G_2}'\sum_{m\in G_1}Q'_{\alpha}\nu_{\alpha\to m}+\sum_{\alpha\in G_2}'\sum_{\alpha'\in
G_2}Q'_{\alpha}\nu_{\alpha\to \alpha'}=&\\
\nonumber & \sum_{\alpha\in G_2}'\sum_{m\in G_1}Q_m\nu_{m\to
\alpha}^{SF}+\sum_{j\in G_3}\sum_{\alpha\in G_2}'P_j\Gamma_{j\to
\alpha} +&\\
\nonumber &\sum_{\alpha\in G_2}'\sum_{\alpha'\in
G_2}Q'_{\alpha'}\nu_{\alpha'\to \alpha}.&
\end{eqnarray}
Following a similar analysis to that above, the second term on the
left hand side is equal to the third term on the right hand side,
and the first term is nonzero only for $m=i$, so one obtains

$
 \sum_{\alpha\in G_2}Q'_\alpha \nu_{\alpha\to
i}=\sum_{m\in G_1}Q_m\sum_{\alpha\in
G_2}'\nu_{m\to\alpha}^{SF}+\sum_{j\in G_3}P_j \sum_{\alpha\in
G_2}'\Gamma_{j\to \alpha}.
$

Substituting this equation and Eq.~\ref{PvsQ} into Eq.~\ref{msb1},
and following several lines of algebra, one obtains the
renormalized masters equation in the space of states $G_1$, with
the renormalized rate

\begin{eqnarray}
\nonumber
 &\Gamma_{m\to i}^{ren}=\nu_{m\to i}^{SF}+\sum_{\alpha\in G_2}^{\nu_{\alpha\to i}\neq 0}\nu_{m\to
 \alpha}^{SF}&\\
 \nonumber
 &+\sum_{j \in G_3}(\Gamma_{m\to j}+\sum_{\beta\in
 G_4}^{\nu_{\beta\to j}\neq 0}\Gamma_{m\to\beta})&\\
 &\times\frac{\Gamma_{j\to i}+\sum_{\alpha\in G_2}^{\nu_{\alpha\to i}\neq 0}\Gamma_{j\to\alpha}}{\sum_{m'\in G_1}\Gamma_{j\to m'}
 +\sum_{\alpha\in G_2}\Gamma_{j\to\alpha}}&\label{gamren}
\end{eqnarray}

This expression is the same as that obtained intuitively in the
main text. The right hand side in the first row is the
renormalized spin-flip rate, $\Omega_{m\to i}=\nu_{m\to
i}^{SF}+\sum_{\alpha\in G_2}^{\nu_{\alpha\to i}\neq 0}\nu_{m\to
 \alpha}^{SF}$. $\nu_{m\to
i}^{SF}$ is the direct spin-flip transition rate and the
 sum is taken over the excited states that can undergo a
 spin-conserving transition into the final state $|i>$.

 In the spin-flip process the spin decreases by $\hbar$, hence $\nu_{m\to
 \i}^{SF}\neq 0$ only if $\Delta S_z=S_z^i-S_z^m=-\hbar$, and
 $\nu_{m\to
 \alpha}^{SF}\neq 0$ only if $\Delta S_z=S_z^\alpha-S_z^m=-\hbar$. It
 follows that $\Omega_{m\to i}\neq 0$ only if $i=m-1$. In that case,
 one can denote
$\Omega_{m\to m-1}=\Omega_m=\nu_{m\to m-1}^{SF}+\sum_{\alpha\in
G_2}^{\nu_{\alpha\to m-1}\neq 0}\nu_{m\to
 \alpha}^{SF}$.
 $\Omega_m$ in this equation is the renormalized spin-flip rate
given by
 Eq.~\ref{spinfliprate}.

Next we examine the second and the third row of Eq.~\ref{gamren}.
The sum over $j$ is taken over intermediate states with an extra
added electron. Writing the rates $\Gamma_{x\to y}$ explicitly
(which is not shown here), it can be seen that the rates of
tunneling in to the various unoccupied single electron states are
multiplied by the rates of tunneling-out from the various occupied
single electron states. The contribution to $\Gamma_{m\to
i}^{ren}$ is nonzero only if $i=m-1$ or $i=m$ or $i=m+1$, because
a sequential tunneling process changes the spin by $-1$ or $0$ or
$1$.

Consider the expression $\Gamma_{m\to m+1}^{ren}$ in
Eq.~\ref{gamren}. In the second row term $\sum_{j \in
G_3}(\Gamma_{m\to j}+\sum_{\beta\in
 G_4}^{\nu_{\beta\to j}\neq 0}\Gamma_{m\to\beta})\times (...)$, $j$ indicates
 the
 state obtained by adding an electron into the
 lowest unoccupied single-electron level with spin up. The
 sum over $\beta$ is taken over the excited states that can relax
 via
 spin-conserving transition into $|j>$. One obtains $\sum_{j \in G_3}(\Gamma_{m\to j}+\sum_{\beta\in
 G_4}^{\nu_{\beta\to j}\neq
 0}\Gamma_{m\to\beta})\times(...)=\sum_{k=m+1}^N\Gamma_k(1+P)\times (...)$,
 where $
\Gamma_k$ is the tunnel-in rate across the left lead into an
unoccupied single-electron energy level $k$. This expression is
the same as that in Eq.~\ref{itoi+1}.

The third row in Eq.~\ref{gamren} can be interpreted as the
probability that the nanoparticle in state $j$ will discharge a
spin-down electron, which is the same as $\pi_i(\downarrow)$ in
Eq.~\ref{itoi+1}. Substituting the tunnel rates into
Eq.~\ref{gamren} explicitly, one obtains that the third row of
Eq.~\ref{gamren} is the same as Eq.~\ref{pidn}.

In summary,  the model of electron transport from the intuitive
approach is derived in this appendix, using the masters equations.

\bibliography{career1}

\begin{thebibliography}{44}
\expandafter\ifx\csname natexlab\endcsname\relax\def\natexlab#1{#1}\fi
\expandafter\ifx\csname bibnamefont\endcsname\relax
  \def\bibnamefont#1{#1}\fi
\expandafter\ifx\csname bibfnamefont\endcsname\relax
  \def\bibfnamefont#1{#1}\fi
\expandafter\ifx\csname citenamefont\endcsname\relax
  \def\citenamefont#1{#1}\fi
\expandafter\ifx\csname url\endcsname\relax
  \def\url#1{\texttt{#1}}\fi
\expandafter\ifx\csname urlprefix\endcsname\relax\def\urlprefix{URL }\fi
\providecommand{\bibinfo}[2]{#2}
\providecommand{\eprint}[2][]{\url{#2}}

\bibitem[{\citenamefont{Seneor et~al.}(2007)\citenamefont{Seneor,
  Bernand-Mantel, and Petroff}}]{seneor}
\bibinfo{author}{\bibfnamefont{P.}~\bibnamefont{Seneor}},
  \bibinfo{author}{\bibfnamefont{A.}~\bibnamefont{Bernand-Mantel}},
  \bibnamefont{and} \bibinfo{author}{\bibfnamefont{F.}~\bibnamefont{Petroff}},
  \bibinfo{journal}{J. Phys.: Condens. Matter} \textbf{\bibinfo{volume}{99}},
  \bibinfo{pages}{165222} (\bibinfo{year}{2007}).

\bibitem[{\citenamefont{Ernult et~al.}(2007)\citenamefont{Ernult, Yakushiji,
  Mitani, and Takanashi}}]{ernult}
\bibinfo{author}{\bibfnamefont{F.}~\bibnamefont{Ernult}},
  \bibinfo{author}{\bibfnamefont{K.}~\bibnamefont{Yakushiji}},
  \bibinfo{author}{\bibfnamefont{S.}~\bibnamefont{Mitani}}, \bibnamefont{and}
  \bibinfo{author}{\bibfnamefont{K.}~\bibnamefont{Takanashi}},
  \bibinfo{journal}{J. Phys. Cond. Mat.} \textbf{\bibinfo{volume}{19}},
  \bibinfo{pages}{1652140} (\bibinfo{year}{2007}).

\bibitem[{\citenamefont{Johnson and Silsbee}(1985)}]{johnson}
\bibinfo{author}{\bibfnamefont{M.}~\bibnamefont{Johnson}} \bibnamefont{and}
  \bibinfo{author}{\bibfnamefont{R.~H.} \bibnamefont{Silsbee}},
  \bibinfo{journal}{Phys. Rev. Lett.} \textbf{\bibinfo{volume}{55}},
  \bibinfo{pages}{1790} (\bibinfo{year}{1985}).

\bibitem[{\citenamefont{Jedema et~al.}(2001)\citenamefont{Jedema, Filip, and
  van Wees}}]{jedema}
\bibinfo{author}{\bibfnamefont{F.~J.} \bibnamefont{Jedema}},
  \bibinfo{author}{\bibfnamefont{A.~T.} \bibnamefont{Filip}}, \bibnamefont{and}
  \bibinfo{author}{\bibfnamefont{B.~J.} \bibnamefont{van Wees}},
  \bibinfo{journal}{Nature} \textbf{\bibinfo{volume}{410}},
  \bibinfo{pages}{345} (\bibinfo{year}{2001}).

\bibitem[{\citenamefont{Bernand-Mantel
  et~al.}(2006)\citenamefont{Bernand-Mantel, Seneor, Lidgi, Munoz, Cros, Fusil,
  Bouzehouane, Deranlot, Vaures, Petroff et~al.}}]{bernard}
\bibinfo{author}{\bibfnamefont{A.}~\bibnamefont{Bernand-Mantel}},
  \bibinfo{author}{\bibfnamefont{P.}~\bibnamefont{Seneor}},
  \bibinfo{author}{\bibfnamefont{N.}~\bibnamefont{Lidgi}},
  \bibinfo{author}{\bibfnamefont{M.}~\bibnamefont{Munoz}},
  \bibinfo{author}{\bibfnamefont{V.}~\bibnamefont{Cros}},
  \bibinfo{author}{\bibfnamefont{S.}~\bibnamefont{Fusil}},
  \bibinfo{author}{\bibfnamefont{K.}~\bibnamefont{Bouzehouane}},
  \bibinfo{author}{\bibfnamefont{C.}~\bibnamefont{Deranlot}},
  \bibinfo{author}{\bibfnamefont{A.}~\bibnamefont{Vaures}},
  \bibinfo{author}{\bibfnamefont{F.}~\bibnamefont{Petroff}},
  \bibnamefont{et~al.}, \bibinfo{journal}{Appl. Phys. Lett.}
  \textbf{\bibinfo{volume}{89}}, \bibinfo{pages}{062502}
  (\bibinfo{year}{2006}).

\bibitem[{\citenamefont{Wei et~al.}(2007)\citenamefont{Wei, Malec, and
  Davidovic}}]{wei}
\bibinfo{author}{\bibfnamefont{Y.~G.} \bibnamefont{Wei}},
  \bibinfo{author}{\bibfnamefont{C.~E.} \bibnamefont{Malec}}, \bibnamefont{and}
  \bibinfo{author}{\bibfnamefont{D.}~\bibnamefont{Davidovic}},
  \bibinfo{journal}{Phys. Rev. B} \textbf{\bibinfo{volume}{76}},
  \bibinfo{pages}{195327} (\bibinfo{year}{2007}).

\bibitem[{\citenamefont{Barnas and Fert}(1998{\natexlab{a}})}]{barnas}
\bibinfo{author}{\bibfnamefont{J.}~\bibnamefont{Barnas}} \bibnamefont{and}
  \bibinfo{author}{\bibfnamefont{A.}~\bibnamefont{Fert}},
  \bibinfo{journal}{Phys. Rev. Lett.} \textbf{\bibinfo{volume}{80}},
  \bibinfo{pages}{1058} (\bibinfo{year}{1998}{\natexlab{a}}).

\bibitem[{\citenamefont{Majumdar and Hershfield}(1998)}]{majumdar}
\bibinfo{author}{\bibfnamefont{K.}~\bibnamefont{Majumdar}} \bibnamefont{and}
  \bibinfo{author}{\bibfnamefont{S.}~\bibnamefont{Hershfield}},
  \bibinfo{journal}{Phys. Rev. B} \textbf{\bibinfo{volume}{57}},
  \bibinfo{pages}{11521} (\bibinfo{year}{1998}).

\bibitem[{\citenamefont{Barnas and Fert}(1998{\natexlab{b}})}]{barnas1}
\bibinfo{author}{\bibfnamefont{J.}~\bibnamefont{Barnas}} \bibnamefont{and}
  \bibinfo{author}{\bibfnamefont{A.}~\bibnamefont{Fert}},
  \bibinfo{journal}{Europhysics letters} \textbf{\bibinfo{volume}{44}},
  \bibinfo{pages}{85} (\bibinfo{year}{1998}{\natexlab{b}}).

\bibitem[{\citenamefont{Brataas
  et~al.}(1999{\natexlab{a}})\citenamefont{Brataas, Nazarov, Inoue, and
  Bauer}}]{brataas}
\bibinfo{author}{\bibfnamefont{A.}~\bibnamefont{Brataas}},
  \bibinfo{author}{\bibfnamefont{Y.~V.} \bibnamefont{Nazarov}},
  \bibinfo{author}{\bibfnamefont{J.}~\bibnamefont{Inoue}}, \bibnamefont{and}
  \bibinfo{author}{\bibfnamefont{G.~E.~W.} \bibnamefont{Bauer}},
  \bibinfo{journal}{European physical journal B} \textbf{\bibinfo{volume}{9}},
  \bibinfo{pages}{421} (\bibinfo{year}{1999}{\natexlab{a}}).

\bibitem[{\citenamefont{Brataas
  et~al.}(1999{\natexlab{b}})\citenamefont{Brataas, Nazarov, Inoue, and
  Bauer}}]{brataas1}
\bibinfo{author}{\bibfnamefont{A.}~\bibnamefont{Brataas}},
  \bibinfo{author}{\bibfnamefont{Y.~V.} \bibnamefont{Nazarov}},
  \bibinfo{author}{\bibfnamefont{J.}~\bibnamefont{Inoue}}, \bibnamefont{and}
  \bibinfo{author}{\bibfnamefont{G.~E.~W.} \bibnamefont{Bauer}},
  \bibinfo{journal}{Phys. Rev. B} \textbf{\bibinfo{volume}{59}},
  \bibinfo{pages}{93} (\bibinfo{year}{1999}{\natexlab{b}}).

\bibitem[{\citenamefont{Korotkov and Safarov}(1999)}]{korotkov1}
\bibinfo{author}{\bibfnamefont{A.~N.} \bibnamefont{Korotkov}} \bibnamefont{and}
  \bibinfo{author}{\bibfnamefont{V.~I.} \bibnamefont{Safarov}},
  \bibinfo{journal}{Phys. Rev. B} \textbf{\bibinfo{volume}{59}},
  \bibinfo{pages}{89} (\bibinfo{year}{1999}).

\bibitem[{\citenamefont{Barnas and Fert}(1999)}]{barnas3}
\bibinfo{author}{\bibfnamefont{J.}~\bibnamefont{Barnas}} \bibnamefont{and}
  \bibinfo{author}{\bibfnamefont{A.}~\bibnamefont{Fert}}, \bibinfo{journal}{J.
  Magn. Magn. Matter.} \textbf{\bibinfo{volume}{192}}, \bibinfo{pages}{391}
  (\bibinfo{year}{1999}).

\bibitem[{\citenamefont{Barnas et~al.}(2000)\citenamefont{Barnas, Martinek,
  Michalek, Bulka, and Fert}}]{barnas2}
\bibinfo{author}{\bibfnamefont{J.}~\bibnamefont{Barnas}},
  \bibinfo{author}{\bibfnamefont{J.}~\bibnamefont{Martinek}},
  \bibinfo{author}{\bibfnamefont{G.}~\bibnamefont{Michalek}},
  \bibinfo{author}{\bibfnamefont{B.~R.} \bibnamefont{Bulka}}, \bibnamefont{and}
  \bibinfo{author}{\bibfnamefont{A.}~\bibnamefont{Fert}},
  \bibinfo{journal}{Phys. Rev. B} \textbf{\bibinfo{volume}{62}},
  \bibinfo{pages}{12363} (\bibinfo{year}{2000}).

\bibitem[{\citenamefont{Imamura et~al.}(1999)\citenamefont{Imamura, Takahashi,
  and Maekawa}}]{imamura2}
\bibinfo{author}{\bibfnamefont{H.}~\bibnamefont{Imamura}},
  \bibinfo{author}{\bibfnamefont{S.}~\bibnamefont{Takahashi}},
  \bibnamefont{and} \bibinfo{author}{\bibfnamefont{S.}~\bibnamefont{Maekawa}},
  \bibinfo{journal}{Phys. Rev. B} \textbf{\bibinfo{volume}{59}},
  \bibinfo{pages}{6017} (\bibinfo{year}{1999}).

\bibitem[{\citenamefont{Kuo and Chen}(2002)}]{kuo}
\bibinfo{author}{\bibfnamefont{W.}~\bibnamefont{Kuo}} \bibnamefont{and}
  \bibinfo{author}{\bibfnamefont{C.~D.} \bibnamefont{Chen}},
  \bibinfo{journal}{Phys. Rev. B} \textbf{\bibinfo{volume}{65}},
  \bibinfo{pages}{104427} (\bibinfo{year}{2002}).

\bibitem[{\citenamefont{Weymann and Barnas}(2003)}]{weymann1}
\bibinfo{author}{\bibfnamefont{I.}~\bibnamefont{Weymann}} \bibnamefont{and}
  \bibinfo{author}{\bibfnamefont{J.}~\bibnamefont{Barnas}},
  \bibinfo{journal}{Phys. Status Solidi b} \textbf{\bibinfo{volume}{236}},
  \bibinfo{pages}{651} (\bibinfo{year}{2003}).

\bibitem[{\citenamefont{Brataas and Wang}(2001)}]{brataas2}
\bibinfo{author}{\bibfnamefont{A.}~\bibnamefont{Brataas}} \bibnamefont{and}
  \bibinfo{author}{\bibfnamefont{X.~H.} \bibnamefont{Wang}},
  \bibinfo{journal}{Phys. Rev. B} \textbf{\bibinfo{volume}{64}},
  \bibinfo{pages}{104434} (\bibinfo{year}{2001}).

\bibitem[{\citenamefont{Wetzels et~al.}(2006)\citenamefont{Wetzels, Bauer, and
  Grifoni}}]{wetzels}
\bibinfo{author}{\bibfnamefont{W.}~\bibnamefont{Wetzels}},
  \bibinfo{author}{\bibfnamefont{G.~E.~W.} \bibnamefont{Bauer}},
  \bibnamefont{and} \bibinfo{author}{\bibfnamefont{M.}~\bibnamefont{Grifoni}},
  \bibinfo{journal}{Phys. Rev. B} \textbf{\bibinfo{volume}{74}},
  \bibinfo{pages}{224406} (\bibinfo{year}{2006}).

\bibitem[{\citenamefont{Weymann et~al.}(20053)\citenamefont{Weymann, Konig,
  Martinek, Barnas, and Schon}}]{weymann2}
\bibinfo{author}{\bibfnamefont{I.}~\bibnamefont{Weymann}},
  \bibinfo{author}{\bibfnamefont{J.}~\bibnamefont{Konig}},
  \bibinfo{author}{\bibfnamefont{J.}~\bibnamefont{Martinek}},
  \bibinfo{author}{\bibfnamefont{J.}~\bibnamefont{Barnas}}, \bibnamefont{and}
  \bibinfo{author}{\bibfnamefont{G.}~\bibnamefont{Schon}},
  \bibinfo{journal}{Phys. Rev. B} \textbf{\bibinfo{volume}{72}},
  \bibinfo{pages}{115334} (\bibinfo{year}{20053}).

\bibitem[{\citenamefont{Braun et~al.}(2005)\citenamefont{Braun, Konig, and
  Martinek}}]{braun}
\bibinfo{author}{\bibfnamefont{M.}~\bibnamefont{Braun}},
  \bibinfo{author}{\bibfnamefont{J.}~\bibnamefont{Konig}}, \bibnamefont{and}
  \bibinfo{author}{\bibfnamefont{J.}~\bibnamefont{Martinek}},
  \bibinfo{journal}{Europhys. Lett.} \textbf{\bibinfo{volume}{72}},
  \bibinfo{pages}{294} (\bibinfo{year}{2005}).

\bibitem[{\citenamefont{Urban et~al.}(2007)\citenamefont{Urban, Braun, and
  Konig}}]{braun1}
\bibinfo{author}{\bibfnamefont{D.}~\bibnamefont{Urban}},
  \bibinfo{author}{\bibfnamefont{M.}~\bibnamefont{Braun}}, \bibnamefont{and}
  \bibinfo{author}{\bibfnamefont{J.}~\bibnamefont{Konig}},
  \bibinfo{journal}{Phys. Rev. B} \textbf{\bibinfo{volume}{76}},
  \bibinfo{pages}{125306} (\bibinfo{year}{2007}).

\bibitem[{\citenamefont{van~der Molen et~al.}(2006)\citenamefont{van~der Molen,
  Tombros, and van Wees}}]{vandermolen}
\bibinfo{author}{\bibfnamefont{S.~J.} \bibnamefont{van~der Molen}},
  \bibinfo{author}{\bibfnamefont{N.}~\bibnamefont{Tombros}}, \bibnamefont{and}
  \bibinfo{author}{\bibfnamefont{B.~J.} \bibnamefont{van Wees}},
  \bibinfo{journal}{Phys. Rev. B} \textbf{\bibinfo{volume}{73}},
  \bibinfo{pages}{220406(R)} (\bibinfo{year}{2006}).

\bibitem[{\citenamefont{Matveev et~al.}(2000)\citenamefont{Matveev, Glazman,
  and Larkin}}]{matveev}
\bibinfo{author}{\bibfnamefont{K.~A.} \bibnamefont{Matveev}},
  \bibinfo{author}{\bibfnamefont{L.~I.} \bibnamefont{Glazman}},
  \bibnamefont{and} \bibinfo{author}{\bibfnamefont{A.~I.}
  \bibnamefont{Larkin}}, \bibinfo{journal}{Phys. Rev. Lett.}
  \textbf{\bibinfo{volume}{85}}, \bibinfo{pages}{2789} (\bibinfo{year}{2000}).

\bibitem[{\citenamefont{Brouwer et~al.}(2000)\citenamefont{Brouwer, Waintal,
  and Halperin}}]{brouwer}
\bibinfo{author}{\bibfnamefont{P.~W.} \bibnamefont{Brouwer}},
  \bibinfo{author}{\bibfnamefont{X.}~\bibnamefont{Waintal}}, \bibnamefont{and}
  \bibinfo{author}{\bibfnamefont{B.~I.} \bibnamefont{Halperin}},
  \bibinfo{journal}{Phys. Rev. Lett.} \textbf{\bibinfo{volume}{85}},
  \bibinfo{pages}{369} (\bibinfo{year}{2000}).

\bibitem[{\citenamefont{Bonet et~al.}(2002)\citenamefont{Bonet, Deshmukh, and
  Ralph}}]{bonet}
\bibinfo{author}{\bibfnamefont{E.}~\bibnamefont{Bonet}},
  \bibinfo{author}{\bibfnamefont{M.~M.} \bibnamefont{Deshmukh}},
  \bibnamefont{and} \bibinfo{author}{\bibfnamefont{D.~C.} \bibnamefont{Ralph}},
  \bibinfo{journal}{Phys. Rev. B} \textbf{\bibinfo{volume}{65}},
  \bibinfo{pages}{045317} (\bibinfo{year}{2002}).

\bibitem[{\citenamefont{Yakushiji et~al.}(2001)\citenamefont{Yakushiji, Mitani,
  Takanashi, Takahashi, Maekawa, Imamura, and Fujimori}}]{yakushiji}
\bibinfo{author}{\bibfnamefont{K.}~\bibnamefont{Yakushiji}},
  \bibinfo{author}{\bibfnamefont{S.}~\bibnamefont{Mitani}},
  \bibinfo{author}{\bibfnamefont{K.}~\bibnamefont{Takanashi}},
  \bibinfo{author}{\bibfnamefont{S.}~\bibnamefont{Takahashi}},
  \bibinfo{author}{\bibfnamefont{S.}~\bibnamefont{Maekawa}},
  \bibinfo{author}{\bibfnamefont{H.}~\bibnamefont{Imamura}}, \bibnamefont{and}
  \bibinfo{author}{\bibfnamefont{H.}~\bibnamefont{Fujimori}},
  \bibinfo{journal}{Appl. Phys. Lett.} \textbf{\bibinfo{volume}{78}},
  \bibinfo{pages}{515} (\bibinfo{year}{2001}).

\bibitem[{\citenamefont{Zhang et~al.}(2005)\citenamefont{Zhang, Wang, Wei, Liu,
  and Davidovi\'c}}]{zhang1}
\bibinfo{author}{\bibfnamefont{L.}~\bibnamefont{Zhang}},
  \bibinfo{author}{\bibfnamefont{C.}~\bibnamefont{Wang}},
  \bibinfo{author}{\bibfnamefont{Y.}~\bibnamefont{Wei}},
  \bibinfo{author}{\bibfnamefont{X.}~\bibnamefont{Liu}}, \bibnamefont{and}
  \bibinfo{author}{\bibfnamefont{D.}~\bibnamefont{Davidovi\'c}},
  \bibinfo{journal}{Phys. Rev. B} \textbf{\bibinfo{volume}{72}},
  \bibinfo{pages}{155445} (\bibinfo{year}{2005}).

\bibitem[{\citenamefont{Yafet}(1963)}]{yafet}
\bibinfo{author}{\bibfnamefont{Y.}~\bibnamefont{Yafet}}, \bibinfo{journal}{Sol.
  State Phys.} \textbf{\bibinfo{volume}{14}}, \bibinfo{pages}{1}
  (\bibinfo{year}{1963}).

\bibitem[{\citenamefont{Adam et~al.}(2002)\citenamefont{Adam, Polianski,
  Waintal, and Brouwer}}]{adam}
\bibinfo{author}{\bibfnamefont{S.}~\bibnamefont{Adam}},
  \bibinfo{author}{\bibfnamefont{M.~L.} \bibnamefont{Polianski}},
  \bibinfo{author}{\bibfnamefont{X.}~\bibnamefont{Waintal}}, \bibnamefont{and}
  \bibinfo{author}{\bibfnamefont{P.~W.} \bibnamefont{Brouwer}},
  \bibinfo{journal}{Phys. Rev. B} \textbf{\bibinfo{volume}{66}},
  \bibinfo{pages}{195412} (\bibinfo{year}{2002}).

\bibitem[{\citenamefont{Elliot}(1954)}]{elliot}
\bibinfo{author}{\bibfnamefont{R.~J.} \bibnamefont{Elliot}},
  \bibinfo{journal}{Phys. Rev.} \textbf{\bibinfo{volume}{96}},
  \bibinfo{pages}{266} (\bibinfo{year}{1954}).

\bibitem[{\citenamefont{Fabian and Sarma}(1998)}]{fabian}
\bibinfo{author}{\bibfnamefont{J.}~\bibnamefont{Fabian}} \bibnamefont{and}
  \bibinfo{author}{\bibfnamefont{S.~D.} \bibnamefont{Sarma}},
  \bibinfo{journal}{Phys. Rev. Lett.} \textbf{\bibinfo{volume}{81}},
  \bibinfo{pages}{5624} (\bibinfo{year}{1998}).

\bibitem[{\citenamefont{Fabian and Sarma}(1999)}]{fabian2}
\bibinfo{author}{\bibfnamefont{J.}~\bibnamefont{Fabian}} \bibnamefont{and}
  \bibinfo{author}{\bibfnamefont{S.~D.} \bibnamefont{Sarma}},
  \bibinfo{journal}{Phys. Rev. Lett.} \textbf{\bibinfo{volume}{83}},
  \bibinfo{pages}{1211} (\bibinfo{year}{1999}).

\bibitem[{\citenamefont{Jedema et~al.}(2002)\citenamefont{Jedema, Heersche,
  Filip, Baselmans, and van Wees}}]{jedema1}
\bibinfo{author}{\bibfnamefont{F.~J.} \bibnamefont{Jedema}},
  \bibinfo{author}{\bibfnamefont{H.~B.} \bibnamefont{Heersche}},
  \bibinfo{author}{\bibfnamefont{A.~T.} \bibnamefont{Filip}},
  \bibinfo{author}{\bibfnamefont{J.~J.~A.} \bibnamefont{Baselmans}},
  \bibnamefont{and} \bibinfo{author}{\bibfnamefont{B.~J.} \bibnamefont{van
  Wees}}, \bibinfo{journal}{Nature} \textbf{\bibinfo{volume}{416}},
  \bibinfo{pages}{713} (\bibinfo{year}{2002}).

\bibitem[{\citenamefont{Jedema et~al.}(2003)\citenamefont{Jedema, Nijboer,
  Filip, and van Wees}}]{jedema3}
\bibinfo{author}{\bibfnamefont{F.~J.} \bibnamefont{Jedema}},
  \bibinfo{author}{\bibfnamefont{M.~S.} \bibnamefont{Nijboer}},
  \bibinfo{author}{\bibfnamefont{A.~T.} \bibnamefont{Filip}}, \bibnamefont{and}
  \bibinfo{author}{\bibfnamefont{B.~J.} \bibnamefont{van Wees}},
  \bibinfo{journal}{Phys. Rev. B} \textbf{\bibinfo{volume}{67}},
  \bibinfo{pages}{085319} (\bibinfo{year}{2003}).

\bibitem[{\citenamefont{Petta and Ralph}(2001)}]{petta}
\bibinfo{author}{\bibfnamefont{J.~R.} \bibnamefont{Petta}} \bibnamefont{and}
  \bibinfo{author}{\bibfnamefont{D.~C.} \bibnamefont{Ralph}},
  \bibinfo{journal}{Phys. Rev. Lett.} \textbf{\bibinfo{volume}{87}},
  \bibinfo{pages}{266801} (\bibinfo{year}{2001}).

\bibitem[{\citenamefont{Agam et~al.}(1997)\citenamefont{Agam, Wingreen,
  Altshuler, Ralph, and Tinkham}}]{agam}
\bibinfo{author}{\bibfnamefont{O.}~\bibnamefont{Agam}},
  \bibinfo{author}{\bibfnamefont{N.~S.} \bibnamefont{Wingreen}},
  \bibinfo{author}{\bibfnamefont{B.~L.} \bibnamefont{Altshuler}},
  \bibinfo{author}{\bibfnamefont{D.~C.} \bibnamefont{Ralph}}, \bibnamefont{and}
  \bibinfo{author}{\bibfnamefont{M.}~\bibnamefont{Tinkham}},
  \bibinfo{journal}{Phys. Rev. Lett.} \textbf{\bibinfo{volume}{78}},
  \bibinfo{pages}{1956} (\bibinfo{year}{1997}).

\bibitem[{\citenamefont{Ralph et~al.}(1995)\citenamefont{Ralph, Black, and
  Tinkham}}]{ralph}
\bibinfo{author}{\bibfnamefont{D.~C.} \bibnamefont{Ralph}},
  \bibinfo{author}{\bibfnamefont{C.~T.} \bibnamefont{Black}}, \bibnamefont{and}
  \bibinfo{author}{\bibfnamefont{M.}~\bibnamefont{Tinkham}},
  \bibinfo{journal}{Phys. Rev. Lett.} \textbf{\bibinfo{volume}{74}},
  \bibinfo{pages}{3241} (\bibinfo{year}{1995}).

\bibitem[{\citenamefont{Khaetskii and Nazarov}(2000)}]{khaetskii}
\bibinfo{author}{\bibfnamefont{A.~V.} \bibnamefont{Khaetskii}}
  \bibnamefont{and} \bibinfo{author}{\bibfnamefont{Y.~V.}
  \bibnamefont{Nazarov}}, \bibinfo{journal}{Phys. Rev. B}
  \textbf{\bibinfo{volume}{61}}, \bibinfo{pages}{12639} (\bibinfo{year}{2000}).

\bibitem[{\citenamefont{Khaetskii and Nazarov}(2001)}]{khaetskii1}
\bibinfo{author}{\bibfnamefont{A.~V.} \bibnamefont{Khaetskii}}
  \bibnamefont{and} \bibinfo{author}{\bibfnamefont{Y.~V.}
  \bibnamefont{Nazarov}}, \bibinfo{journal}{Phys. Rev. B}
  \textbf{\bibinfo{volume}{64}}, \bibinfo{pages}{125316}
  (\bibinfo{year}{2001}).

\bibitem[{\citenamefont{Hanson et~al.}(2003)\citenamefont{Hanson, Witkamp,
  Vandersypen, van Beveren, Elzerman, and Kouwenhoven}}]{hanson}
\bibinfo{author}{\bibfnamefont{R.}~\bibnamefont{Hanson}},
  \bibinfo{author}{\bibfnamefont{B.}~\bibnamefont{Witkamp}},
  \bibinfo{author}{\bibfnamefont{L.}~\bibnamefont{Vandersypen}},
  \bibinfo{author}{\bibfnamefont{L.~W.} \bibnamefont{van Beveren}},
  \bibinfo{author}{\bibfnamefont{J.}~\bibnamefont{Elzerman}}, \bibnamefont{and}
  \bibinfo{author}{\bibfnamefont{L.}~\bibnamefont{Kouwenhoven}},
  \bibinfo{journal}{Phys. Rev. Lett.} \textbf{\bibinfo{volume}{91}}
  (\bibinfo{year}{2003}).

\bibitem[{\citenamefont{Averin et~al.}(1999)\citenamefont{Averin, Korotkov, and
  Likharev}}]{averin1}
\bibinfo{author}{\bibfnamefont{D.~V.} \bibnamefont{Averin}},
  \bibinfo{author}{\bibfnamefont{A.~N.} \bibnamefont{Korotkov}},
  \bibnamefont{and} \bibinfo{author}{\bibfnamefont{K.~K.}
  \bibnamefont{Likharev}}, \bibinfo{journal}{Phys. Rev. B}
  \textbf{\bibinfo{volume}{44}}, \bibinfo{pages}{6199} (\bibinfo{year}{1999}).

\bibitem[{\citenamefont{Beenakker}(1991)}]{beenakker1}
\bibinfo{author}{\bibfnamefont{C.~W.~J.} \bibnamefont{Beenakker}},
  \bibinfo{journal}{Phys. Rev. B} \textbf{\bibinfo{volume}{44}},
  \bibinfo{pages}{1646} (\bibinfo{year}{1991}).

\bibitem[{\citenamefont{Julliere}(1975)}]{julliere}
\bibinfo{author}{\bibfnamefont{M.}~\bibnamefont{Julliere}},
  \bibinfo{journal}{Phys. Lett. A} \textbf{\bibinfo{volume}{54}},
  \bibinfo{pages}{225} (\bibinfo{year}{1975}).

\end{thebibliography}
\end{document}